\documentclass[aps,prc,preprint,amsmath,amssymb,preprintnumbers,superscriptaddress]{revtex4}

\usepackage{amsmath}
\usepackage{graphicx}
\usepackage{bm}
\usepackage{hyperref}
\usepackage{braket}
\usepackage{multirow}
\usepackage{mathrsfs}
\usepackage{longtable}
\usepackage{pdflscape}

\allowdisplaybreaks[4]

\begin{document}

\title{Deformed relativistic Hartree-Bogoliubov theory in continuum with a point-coupling functional. II. Examples of odd Nd isotopes}

\author{Cong Pan}
\affiliation{State Key Laboratory of Nuclear Physics and Technology, School of Physics, Peking University, Beijing 100871, China}

\author{Myung-Ki Cheoun}
\affiliation{Department of Physics and Origin of Matter and Evolution of Galaxy (OMEG) Institute, Soongsil University, Seoul 156-743, Korea}

\author{Yong-Beom Choi}
\affiliation{Department of Physics, Pusan National University, Busan 46241, Korea}

\author{Jianmin Dong}
\affiliation{Institute of Modern Physics, Chinese Academy of Sciences, Lanzhou 730000, China}
\affiliation{School of Physics, University of Chinese Academy of Sciences, Beijing 100049, China}

\author{Xiaokai Du}
\affiliation{State Key Laboratory of Nuclear Physics and Technology, School of Physics, Peking University, Beijing 100871, China}

\author{Xiao-Hua Fan}
\affiliation{School of Physical Science and Technology, Southwest University, Chongqing 400715, China}

\author{Wei Gao}
\affiliation{School of Physics and Microelectronics, Zhengzhou University, Zhengzhou 450001, China}

\author{Lisheng Geng}
\affiliation{School of Physics, Beihang University, Beijing 102206, China}
\affiliation{School of Physics and Microelectronics, Zhengzhou University, Zhengzhou 450001, China}

\author{Eunja Ha}
\affiliation{Department of Physics, Hanyang University, Seoul, 04763, Korea}

\author{Xiao-Tao He}
\affiliation{College of Materials Science and Technology, Nanjing University of Aeronautics and Astronautics, Nanjing 210016, China}

\author{Jinke Huang}
\affiliation{School of Physics and Microelectronics, Zhengzhou University, Zhengzhou 450001, China}

\author{Kun Huang}
\affiliation{College of Materials Science and Technology, Nanjing University of Aeronautics and Astronautics, Nanjing 210016, China}

\author{Seonghyun Kim}
\affiliation{Department of Physics and Origin of Matter and Evolution of Galaxy (OMEG) Institute, Soongsil University, Seoul 156-743, Korea}

\author{Youngman Kim}
\affiliation{Rare Isotope Science Project, Institute for Basic Science, Daejeon 34000, Korea}

\author{Chang-Hwan Lee}
\affiliation{Department of Physics, Pusan National University, Busan 46241, Korea}

\author{Jenny Lee}
\affiliation{Department of Physics, The University of Hong Kong, Pokfulam 999077, Hong Kong, China}

\author{Zhipan Li}
\affiliation{School of Physical Science and Technology, Southwest University, Chongqing 400715, China}

\author{Zhi-Rui Liu}
\affiliation{College of Materials Science and Technology, Nanjing University of Aeronautics and Astronautics, Nanjing 210016, China}

\author{Yiming Ma}
\affiliation{School of Mathematical Sciences, Peking University, Beijing 100871, China}

\author{Jie Meng} \email{mengj@pku.edu.cn}
\affiliation{State Key Laboratory of Nuclear Physics and Technology, School of Physics, Peking University, Beijing 100871, China}

\author{Myeong-Hwan Mun}
\affiliation{Department of Physics and Origin of Matter and Evolution of Galaxy (OMEG) Institute, Soongsil University, Seoul 156-743, Korea}
\affiliation{Korea Institute of Science and Technology Information, Daejeon 34141, Korea}

\author{Zhongming Niu}
\affiliation{School of Physics and Optoelectronics Engineering, Anhui University, Hefei 230601, China}

\author{Panagiota Papakonstantinou}
\affiliation{Rare Isotope Science Project, Institute for Basic Science, Daejeon 34000, Korea}

\author{Xinle Shang}
\affiliation{Institute of Modern Physics, Chinese Academy of Sciences, Lanzhou 730000, China}
\affiliation{School of Physics, University of Chinese Academy of Sciences, Beijing 100049, China}

\author{Caiwan Shen}
\affiliation{School of Science, Huzhou University, Huzhou 313000, China}

\author{Guofang Shen}
\affiliation{School of Physics, Beihang University, Beijing 102206, China}

\author{Wei Sun}
\affiliation{School of Physical Science and Technology, Southwest University, Chongqing 400715, China}

\author{Xiang-Xiang Sun}
\affiliation{School of Nuclear Science and Technology, University of Chinese Academy of Sciences, Beijing 100049, China}
\affiliation{CAS Key Laboratory of Theoretical Physics, Institute of Theoretical Physics, Chinese Academy of Sciences, Beijing 100190, China}

\author{Jiawei Wu}
\affiliation{College of Materials Science and Technology, Nanjing University of Aeronautics and Astronautics, Nanjing 210016, China}

\author{Xinhui Wu}
\affiliation{State Key Laboratory of Nuclear Physics and Technology, School of Physics, Peking University, Beijing 100871, China}

\author{Xuewei Xia}
\affiliation{School of Physics and Electronic Engineering, Center for Computational Sciences, Sichuan Normal University, Chengdu 610068, China}

\author{Yijun Yan}
\affiliation{Institute of Modern Physics, Chinese Academy of Sciences, Lanzhou 730000, China}
\affiliation{School of Physics, University of Chinese Academy of Sciences, Beijing 100049, China}

\author{To Chung Yiu}
\affiliation{Department of Physics, The University of Hong Kong, Pokfulam 999077, Hong Kong, China}

\author{Kaiyuan Zhang}
\affiliation{State Key Laboratory of Nuclear Physics and Technology, School of Physics, Peking University, Beijing 100871, China}

\author{Shuangquan Zhang}
\affiliation{State Key Laboratory of Nuclear Physics and Technology, School of Physics, Peking University, Beijing 100871, China}

\author{Wei Zhang}
\affiliation{School of Physics and Microelectronics, Zhengzhou University, Zhengzhou 450001, China}

\author{Xiaoyan Zhang}
\affiliation{School of Physics and Optoelectronics Engineering, Anhui University, Hefei 230601, China}

\author{Qiang Zhao}
\affiliation{Center for Exotic Nuclei Studies, Institute for Basic Science, Daejeon 34126, Korea}
\affiliation{State Key Laboratory of Nuclear Physics and Technology, School of Physics, Peking University, Beijing 100871, China}

\author{Ruyou Zheng}
\affiliation{School of Physics, Beihang University, Beijing 102206, China}

\author{Shan-Gui Zhou}
\affiliation{CAS Key Laboratory of Theoretical Physics, Institute of Theoretical Physics, Chinese Academy of Sciences, Beijing 100190, China}
\affiliation{School of Physical Sciences, University of Chinese Academy of Sciences, Beijing 100049, China}
\affiliation{Center of Theoretical Nuclear Physics, National Laboratory of Heavy Ion Accelerator, Lanzhou 730000, China}
\affiliation{Synergetic Innovation Center for Quantum Effects and Application, Hunan Normal University, Changsha 410081, China}

\collaboration{DRHBc Mass Table Collaboration}

\begin{abstract}

\noindent \textbf{Background:}
One fascinating frontier in nuclear physics is the study of exotic nuclei.
The deformed relativistic Hartree-Bogoliubov theory in continuum (DRHBc), which simultaneously includes the nuclear superfluidity, deformation, and continuum effects, can provide proper descriptions for both stable and exotic nuclei.
In Zhang, \textit{et al}. \href{https://doi.org/10.1103/PhysRevC.102.024314}{[Phys. Rev. C 102, 024314 (2020)]}, the DRHBc theory based on the point-coupling density functionals was developed and the DRHBc calculation, previously accessible only for light nuclei, was extended for all even-even nuclei in the nuclear chart.
The ground-state properties for the even-even nuclei with $8 \leq Z \leq 120$ from the DRHBc calculations have been summarized in Zhang, \textit{et al}. \href{https://doi.org/10.1016/j.adt.2022.101488}{[At. Data Nucl. Data Tables 144, 101488 (2022)]}.

\noindent \textbf{Purpose:}
The aim of this work is to extend the point-coupling DRHBc theory to odd-$A$ and odd-odd nuclei and examine its applicability by taking odd-$A$ Nd isotopes as examples.

\noindent \textbf{Method:}
In the DRHBc theory, the densities and potentials with axial deformation are expanded in terms of Legendre polynomials, and the relativistic Hartree-Bogoliubov equations are solved in a Dirac Woods-Saxon basis to include the continuum effects.
For an odd-$A$ or odd-odd nucleus, the blocking effect of unpaired nucleon(s) is taken into account with the equal filling approximation.
To determine its ground state, an automatic blocking procedure is adopted, in which the orbital with the lowest quasiparticle energy is blocked during the iteration.
This procedure is justified by comparing with the results from the orbital-fixed blocking calculations, in which the blocked orbital near the Fermi surface is fixed during the iteration.
The ground states for both light and heavy nuclei can be provided by the automatic blocking procedure as the orbital-fixed blocking procedure, but with considerably reduced computational cost.

\noindent \textbf{Results:}
The numerical details for even-even nuclei, including the convergence on the energy cutoff, angular momentum cutoff and Legendre expansion, are found to be valid for odd-$A$ and odd-odd nuclei as well.
The ground-state properties of odd-$A$ Nd isotopes are calculated with the density functional PC-PK1.
The calculated physical observables, such as binding energies, two-neutron and one-neutron separation energies, and charge radii, are in good agreement with the available experimental data for the whole Nd isotopic chain.

\noindent \textbf{Conclusions:}
The point-coupling DRHBc theory is extended to odd-$A$ and odd-odd nuclei by including the blocking effect.
Taking Nd isotopes including both even-even and odd-$A$ ones as examples, the calculated ground-state properties with PC-PK1 are in good agreement with the available experimental data.
This work paves the way to construct the DRHBc mass table including all even-even, odd-$A$ and odd-odd nuclei in the nuclear chart.

\end{abstract}

\date{\today}

\maketitle

\section{Introduction}

In the previous paper \cite{Zhang2020PRC}, the deformed relativistic Hartree-Bogoliubov theory in continuum (DRHBc) \cite{Zhou2010PRC} based on the point-coupling density functionals was developed for even-even nuclei, and the strategy and techniques to construct a mass table with the deformation and continuum effects were discussed by taking even-even neodymium isotopes as examples.
The aim of this paper is to extend the DRHBc theory based on the point-coupling density functionals to describe the odd-$A$ and odd-odd nuclei in order to construct a complete DRHBc mass table, which will provide valuable predictions for the exotic nuclei beyond the current experimental capability \cite{Thoennessen2013RPP,nndc,Kondev2021CPC,Huang2021CPC,Wang2021CPC}.

The nonrelativistic density functional theory or the covariant density functional theory (CDFT), can provide a self-consistent description for nuclei all over the nuclear chart \cite{Ring1996PPNP,Bender2003RMP,Meng2016book}.
Many systematic investigations on the whole nuclear landscape, especially nuclear masses, have been performed \cite{Erler2012Nat,Samyn2002NPA,Stoitsov2003PRC,Goriely2009PRL_Skyrme,Goriely2013PRC,
Hilaire2007EPJA,Goriely2009PRL_Gogny,Delaroche2010PRC,
Lalazissis1999ADNDT,Geng2005PTP,Meng2013FoP,Zhang2014FoP,Agbemava2014PRC,Afanasjev2015PRC,Lu2015PRC,Pena-Arteaga2016EPJA,Xia2018ADNDT,Yang2021PRC}. 

For exotic nuclei where the Fermi energy is close to the continuum threshold, the pairing interaction can scatter nucleons from bound states to the resonant states in the continuum.
This might lead to a more diffuse density and influence the drip-line location, which is the so-called continuum effect \cite{Meng2006PPNP}.
Based on the CDFT and taking into account pairing correlations and continuum effect, the relativistic continuum Hartree-Bogoliubov (RCHB) theory was developed \cite{Meng1996PRL,Meng1998NPA}, and has achieved great success in descriptions for both stable and exotic nuclei \cite{Meng1998PRL,Meng1998PLB,Meng2002PLB,Meng2002PRC,Zhang2002CPL,Lv2003EPJA,Zhang2005NPA,Meng2006PPNP,Lim2016PRC,Zhang2016CPC}.
In Ref.~\cite{Xia2018ADNDT}, the first nuclear mass table including continuum effect was constructed based on the RCHB theory, and the continuum effects on the limits of the nuclear landscape were studied.

Except for doubly-magic nuclei, most nuclei in the nuclear chart deviate from spherical shape.
Inheriting the advantages of the RCHB theory and including the deformation degree of freedom, the deformed relativistic Hartree-Bogoliubov theory in continuum (DRHBc) was developed in Refs.~\cite{Zhou2010PRC,Li2012PRC}, with the deformed relativistic Hartree-Bogoliubov equations solved in a Dirac Woods-Saxon basis \cite{Zhou2003PRC}.
The DRHBc theory has been successfully applied in many studies on exotic nuclei in the light-mass region \cite{Chen2012PRC,Sun2018PLB,Zhang2019PRC,Sun2020NPA,In2021IJMPE,Yang2021PRL,Sun2021PRC,Sun2021SciB,Sun2021PRC_AMP}. 

In Ref. \cite{Zhang2020PRC}, the DRHBc theory based on the point-coupling functionals was developed for even-even nuclei, and the strategy and techniques to construct a mass table with the deformation and continuum effects by using the DRHBc theory were discussed.
The convergence of the nuclear potentials and densities on the Legendre expansion can be found in Ref.~\cite{Pan2019IJMPE}.
Along this line, interesting topics discussed include the deformation effects on the location of neutron drip line \cite{In2021IJMPE}, the peninsulas of stability beyond the two-neutron drip line \cite{Zhang2021PRC,Pan2021PRC,He2021CPC}, the dynamical correlation energy with a two-dimensional collective Hamiltonian \cite{Sun2022CPC}, and the bubble structure and shape coexistence \cite{Choi2022PRC,Kim2022PRC}.
The DRHBc mass table for even-even nuclei is provided in Ref.~\cite{Zhang2022ADNDT}, where 2583 even-even nuclei with $8\leq Z \leq 120$ are predicted to be bound, and the deformation and continuum effects on the nuclear landscape are investigated.

For odd-$A$ and odd-odd nuclei, the blocking effect of the unpaired nucleon(s) should be considered \cite{Ring1980NMBP}, which may play an important role in the halo structure \cite{Nakada2018PRC,Sun2020NPA,Kasuya2020PTEP}.
In Ref.~\cite{Li2012CPL}, the DRHBc theory based on the meson-exchange functional has been extended to incorporate the blocking effect of odd nucleon(s).

For odd-$A$ and odd-odd nuclei, a challenge for determining the ground state is to block the correct orbital(s).
To this end, the calculation should be done by respectively blocking the corresponding orbitals near the Fermi surface and the result with the lowest energy should be identified as the ground state \cite{Xia2018ADNDT,Zhang2020PRC}.
Due to the deformation degree of freedom, such a computational procedure is quite demanding in the DRHBc calculations.
An optimized procedure to look for the ground state with the correctly blocked orbital is needed.

In this work, the DRHBc theory including the blocking effect based on the point-coupling functional is extended to describe the odd-$A$ and odd-odd nuclei.
The theoretical framework is presented in Section \ref{sec:th}.
The blocking procedure and the numerical details are introduced in Section \ref{sec:num}.
Taking the neodymium isotopic chain as an example, the DRHBc calculated results are presented and compared with the RCHB mass table \cite{Xia2018ADNDT} and the available data \cite{Angeli2013ADNDT,Pritychenko2016ADNDT,Wang2021CPC} in Section \ref{sec:Nd}.
A summary is given in Section \ref{sec:summary}.

\section{Theoretical framework}
\label{sec:th}

The details of the DRHBc theory can be found in Refs. \cite{Zhou2010PRC,Li2012PRC,Li2012CPL,Zhang2020PRC}.
Here a brief introduction is presented in Section \ref{subsec:DRHBc}, and the DRHBc theory for odd-$A$ or odd-odd nucleus is introduced in Section \ref{subsec:odd}.

\subsection{Brief introduction of the DRHBc theory}
\label{subsec:DRHBc}

In the framework of the Hartree-Fock-Bogoliubov (HFB) or relativistic Hartree-Bogoliubov (RHB) theory, the ground state $\ket{\Phi}$ is constructed as \cite{Ring1980NMBP,Meng2016book}
\begin{equation}
	\label{eq:Phi}
	\ket{\Phi} = \prod_k \beta_k \ket{0},
\end{equation}
where $\ket{0}$ is the bare vacuum, $\beta_k$ is the quasiparticle annihilation operator, and $k$ is the quasiparticle index.
The quasiparticle operators $\beta_k^\dagger, \beta_k$ are defined by the unitary Bogoliubov transformation from the particle operators $c_l^\dagger, c_l$ of an arbitrary complete and orthogonal basis, e.g., a harmonic oscillator basis or Woods-Saxon basis,
\begin{equation}
	\label{eq:betak}
	\beta_k^\dagger = \sum_l (U_{lk} c_l^\dagger + V_{lk} c_l) ,
\end{equation}
where the coefficients $U_{lk}$ and $V_{lk}$ are quasiparticle wavefunctions.

The density matrix $\rho$ and pairing tensor $\kappa$,
\begin{align}
	\rho & = V^* V^T, \label{eq:rho} \\
	\kappa & = V^* U^T, \label{eq:kappa}
\end{align}
are the quantities which determine $\ket{\Phi}$ uniquely \cite{Ring1980NMBP}.

In the RHB framework, the quasiparticle wavefunctions $U$ and $V$ are determined by the RHB equations \cite{Kucharek1991ZPA},
\begin{equation}
	\label{eq:RHBeq}
	\left( \begin{matrix}
		\hat h_D - \lambda_\tau & \hat \Delta \\
		-\hat \Delta^* & -\hat h_D^* + \lambda_\tau
		\end{matrix} \right)
	\left( \begin{matrix} U_k \\ V_k \end{matrix} \right) = E_k
	\left( \begin{matrix} U_k \\ V_k \end{matrix} \right) ,
\end{equation}
where $\hat{h}_D$ is the Dirac Hamiltonian, $\lambda_\tau$ is the Fermi energy for neutron or proton ($\tau=n,p$), $\hat{\Delta}$ is the pairing potential, and $E_k$ is the quasiparticle energy.

In coordinate space, the Dirac Hamiltonian is
\begin{equation}
	h_D(\bm{r}) = \bm{\alpha} \cdot \bm{p} + V(\bm{r}) + \beta[M+S(\bm{r})] ,
\end{equation}
where $M$ is the nucleon mass, and $S(\bm{r})$ and $V(\bm{r})$ are the scalar and vector potentials.
The Dirac Hamiltonian can be derived from an effective Lagrangian with either the meson-exchange or point-coupling interaction \cite{Meng2016book}.

For the Lagrangian with the point-coupling interaction, the zero-range interaction is used in the scalar-isoscalar, vector-isoscalar, scalar-isovector, and vector-isovector channels, and the medium and finite-range effects are taken into account by including higher-order and derivative terms, respectively \cite{Burvenich2002PRC,Zhao2010PRC}.
The corresponding potentials $S(\bm{r})$ and $V(\bm{r})$ are
\begin{align}
	\label{eq:S}
	S(\bm{r}) & =  \alpha_S \rho_S + \beta_S \rho^2_S + \gamma_S \rho^3_S + \delta_S \Delta\rho_S , \\
	\label{eq:V}
	V(\bm{r}) & = \alpha_V \rho_V + \gamma_V \rho^3_V + \delta_V \Delta\rho_V + e A^0 + \alpha_{TV}\tau_3\rho_3 +\delta_{TV}\tau_3\Delta\rho_3 .
\end{align}
There are 9 coupling constants, with $\alpha$ corresponding to the four-fermion terms, $\beta$ and $\gamma$ respectively to the third- and fourth-order terms, and $\delta$ to the derivative couplings.
The subscripts $S$, $V$, and $T$ indicate the symmetries of the couplings, i.e., scalar, vector, and isovector, respectively.

The local densities $\rho_S$, $\rho_V$, and $\rho_3$ are defined as
\begin{align}
	\rho_S(\bm{r}) & = \sum_{k>0} V_k^\dagger(\bm{r}) \gamma_0 V_k (\bm{r}) , \\
	\rho_V(\bm{r}) & = \sum_{k>0} V_k^\dagger(\bm{r})  V_k (\bm{r}) , \\
	\rho_3(\bm{r}) & = \sum_{k>0} V_k^\dagger(\bm{r}) \tau_3 V_k (\bm{r}) .
\end{align}
Here the no-sea approximation is adopted, i.e., the summation runs over the quasiparticle states in the Fermi sea.

The pairing potential is
\begin{equation}
	\Delta(\bm{r}_1,\bm{r}_2) = V^{pp}(\bm{r}_1,\bm{r}_2) \kappa(\bm{r}_1,\bm{r}_2) ,
\end{equation}
where for simplicity the spin and isospin degrees of freedom are not shown, $\kappa$ is the pairing tensor \cite{Ring1980NMBP}, and $V^{pp}$ is a density-dependent zero-range pairing force,
\begin{equation}
	V^{pp} (\bm{r}_1,\bm{r}_2) = V_0 \frac{1}{2} (1-P^\sigma) \delta(\bm{r}_1 - \bm{r}_2) \left( 1-\frac{\rho(\bm r_1)}{\rho_{\mathrm{sat}}} \right) ,
\end{equation}
where $V_0$ is the pairing strength, $\rho_{\mathrm{sat}}$ is the saturation density of nuclear matter, and $(1-P^\sigma)/2$ is the projector for the spin $S=0$ component in the pairing channel.

For an axially deformed nucleus with spatial reflection symmetry, the potentials and densities can be expanded in terms of Legendre polynomials,
\begin{equation}
	\label{eq:Legexp}
	f(\bm{r}) = \sum_\lambda f_\lambda(r) P_\lambda(\cos\theta), \quad \lambda=0,2,4,\dots, \lambda_{\max}
\end{equation}
with
\begin{equation}
	f_\lambda(r) = \frac{2\lambda+1}{4\pi} \int d\Omega f(\bm{r}) P_\lambda(\Omega).
\end{equation}
In order to take into account the continuum effect, in the DRHBc theory \cite{Zhou2010PRC,Li2012PRC}, the RHB equations \eqref{eq:RHBeq} are solved in a spherical Dirac Woods-Saxon (WS) basis \cite{Zhou2003PRC}.
Due to the spatial reflection symmetry and axial symmetry, the full RHB matrix can be  decomposed into blocks characterized by quantum numbers $m^\pi$, where $\pi$ is parity and $m$ is the third component of the angular momentum.
The diagonalization of the RHB matrix yields the quasiparticle wavefunctions, which can be used to construct densities and potentials.

After self-consistently solving the RHB equations \eqref{eq:RHBeq}, the physical observables including the total energy, root-mean-square (rms) radius, and deformation can be calculated.
The total energy of a nucleus is
\begin{equation}
	\label{eq:ERHB}
	\begin{aligned}
		E_{\mathrm{RHB}} = & E_{\mathrm{nucleon}} + E_\mathrm{pair} \\
		& - \int \mathrm{d}^3 \bm r \left(\frac{1}{2}\alpha_S \rho_S^2 + \frac{1}{2}\alpha_V \rho_V^2 + \frac{1}{2}\alpha_{TV}\rho_3^2 \right.\\
		& + \left.\frac{2}{3}\beta_S \rho^3_S + \frac{3}{4}\gamma_S \rho^4_S + \frac{3}{4}\gamma_V \rho^4_V + \frac{1}{2}\delta_S \rho_S \Delta\rho_S \right. \\
		& + \left.\frac{1}{2} \delta_V \rho_V \Delta\rho_V +\frac{1}{2}\delta_{TV}\rho_3 \Delta\rho_3 +\frac{1}{2} \rho_p e A^0\right) \\
		& + E_{\mathrm{c.m.}} ,
	\end{aligned}
\end{equation}
where the nucleon energy $E_{\mathrm{nucleon}}$ reads
\begin{equation}
	\label{eq:Enuc}
	E_{\mathrm{nucleon}} = \sum_{k>0} (\lambda - E_k) v_k^2 - 2E_{\mathrm{pair}} ,
\end{equation}
with
\begin{equation}
	v_k^2=\int d^3 \bm{r} V_k^\dag(\bm{r}) V_k(\bm{r}) .
\end{equation}
The pairing energy $E_{\mathrm{pair}}$, with the zero-range pairing force, is calculated by
\begin{equation}
	E_{\mathrm{pair}} = -\frac{1}{2} \int d^3 \bm{r} \kappa(\bm{r}) \Delta(\bm{r}).
\end{equation}
The center-of-mass (c.m.) correction energy is calculated by
\begin{equation}
	E_{\mathrm{c.m.}} = -\frac{ \braket{\hat {\bm{P}}^2} }{2MA} ,
\end{equation}
where $A$ is the mass number, and $\hat{\bm{P}} = \sum_i^A \hat{\bm{p}}_i$ is the total momentum in the c.m. frame \cite{Bender2000EPJA,Long2004PRC,Zhao2009CPL}.

For deformed nuclei, due to the breaking of the rotational symmetry in the mean-field approximation, the energy gained by the restoration of rotational symmetry, i.e., the rotational correction energy $E_{\mathrm{rot}}$, are taken into account by the cranking approximation \cite{Zhao2010PRC,Zhang2020PRC},
\begin{equation}
	\label{eq:Erot}
	E_{\mathrm{rot}} = -\frac{ \braket{\hat{\bm{J}}^2} }{2 \mathscr{I} } ,
\end{equation}
where $\hat{\bm{J}}$ is the total angular momentum operator, and $\mathscr{I}$ is the moment of inertia obtained with the Inglis-Belyaev formula \cite{Ring1980NMBP}.

The rms radius is calculated by
\begin{equation}
	R_{\tau} = \braket{r^2}^{1/2} = \sqrt{ \frac{1}{N_\tau} \int d^3 \bm{r} \left[  r^2 \rho_\tau(\bm{r}) \right] } ,
\end{equation}
where $\tau$ represents the neutron, proton, or nucleon, $\rho_\tau$ is the corresponding vector density, and $N_\tau$ refers to the corresponding particle number.
The rms charge radii is calculated by
\begin{equation}
	R_{\mathrm{ch}}= \sqrt{R_{p}^2 + 0.64 ~ \mathrm{fm}^2}.
\end{equation}
The quadrupole deformation is calculated by
\begin{equation}
	\beta_{\tau,2} = \frac{\sqrt{5\pi}Q_{\tau,2}}{3N_\tau \braket{ r_\tau^2 } } ,
\end{equation}
where $Q_{\tau,2}$ is the intrinsic quadrupole moment
\begin{equation}
	Q_{\tau,2} = \sqrt{\frac{16\pi}{5}} \braket{r^2 Y_{20}(\theta,\varphi)} .
\end{equation}

The canonical basis $\ket{\psi_i}$ can be obtained by diagonalizing the density matrix $\rho$ \cite{Ring1980NMBP},
\begin{equation}
	\rho \ket{\psi_i} = v_i^2 \ket{\psi_i} ,
\end{equation}
where $\rho$ is given in Eq.~\eqref{eq:rho}, and the eigenvalue $v_i^2$ is the corresponding occupation probability of $\ket{\psi_i}$.

\subsection{The DRHBc theory for odd-$A$ or odd-odd nucleus}
\label{subsec:odd}

For an odd-$A$ or odd-odd nucleus, the blocking effect of the unpaired nucleon(s) needs to be considered \cite{Ring1980NMBP}.
Starting from the ground state of an even-even system $\ket{\Phi}$ as defined in Eq.~\eqref{eq:Phi}, the ground state for a system with an unpaired particle can be described by a one-quasiparticle state
\begin{equation}
	\label{eq:Phi1}
	\ket{\Phi_1} = \beta_{k_b}^\dagger \ket{\Phi} = \beta_{k_b}^\dagger \prod_k \beta_k \ket{0},
\end{equation}
where $\beta_{k_b}^\dagger$ corresponds to the quasiparticle state properly blocked.
In other words, the one-quasiparticle state $\ket{\Phi_1}$ is the vacuum with respect to the set of quasiparticle operators $(\beta_1, ~ \dots, ~ \beta_{k_b}^\dagger, ~ \dots, ~ \beta_N)$.
That is, the blocking effect can be realized by the exchange of $\beta_{k_b} \leftrightarrow \beta_{k_b}^\dagger$.
According to Eqs.~\eqref{eq:betak} and \eqref{eq:RHBeq}, this exchange corresponds to the exchange of the columns $(V_{k_b}^*, U_{k_b}^*) \leftrightarrow (U_{k_b},V_{k_b})$ and that of the energy $E_{k_b} \leftrightarrow -E_{k_b}$.
Similarly, the blocking effect in a multi-quasiparticle configuration can be treated.

For an axially deformed odd-$A$ nucleus, the blocked orbital $k_b$ breaks the time reversal symmetry and the currents appear.
Due to the axial symmetry, $m$ remains a good quantum number, but the single-particle state with $+m$ and its conjugate state with $-m$ are no longer degenerate.
For simplicity, the equal filling approximation (EFA) is usually adopted~\cite{Perez-Martin2008PRC,Li2012CPL}, where the currents vanish and the two configurations of a particle in the $+m$ space and a particle in the $-m$ space are averaged in a statistical manner.
In this way we obtain in each step of the iteration the fields with the time reversal symmetry.
Correspondingly, the density matrix $\rho$ and pairing tensor $\kappa$ in Eqs.~\eqref{eq:rho} and \eqref{eq:kappa} are replaced by~\cite{Li2012CPL}
\begin{align}
	\label{eq:blkrho}
	\rho' & = \rho + \frac{1}{2} (U_{k_b} U_{k_b}^{*T} - V_{k_b}^* V_{k_b}^{T}), \\
	\label{eq:blkkap}
	\kappa' & = \kappa - \frac{1}{2} (U_{k_b} V_{k_b}^{*T} + V_{k_b}^* U_{k_b}^T).
\end{align}

The total energy of an odd-$A$ nucleus is still given by Eq.~\eqref{eq:ERHB}, but the densities and pairing tensors are replaced respectively by Eqs.~\eqref{eq:blkrho} and \eqref{eq:blkkap}.
Thus the nucleon energy in Eq.~\eqref{eq:Enuc} becomes
\begin{equation}
	\label{eq:Enucodd}
	E_{\mathrm{nucleon}} = 2 \sum_{k>0}^{(m>0)} (\lambda - E_k) v_k^2 + (\lambda + E_{k_b}) u_{k_b}^2 - (\lambda - E_{k_b}) v_{k_b}^2 - 2E_{\mathrm{pair}} ,
\end{equation}
where $u_k^2 = 1-v_k^2$.
Note that due to the assumed time-reversal symmetry, the contributions from the orbitals with $m>0$ and those with $m<0$ are the same, and thus one only needs to consider the positive-$m$ orbitals in Eq.~\eqref{eq:Enucodd}.
For the rotational correction energy \eqref{eq:Erot}, since it is calculated in the canonical basis and the EFA is implemented here, the ground-state wavefunction in the canonical basis of an odd-$A$ nucleus has a form similar to that of an even-even nucleus, and as a result the calculation of $E_{\mathrm{rot}}$ is the same as an even-even one.

\section{Numerical details}
\label{sec:num}

\subsection{Blocking procedure}

To determine the ground state of an odd-$A$ or odd-odd nucleus, one needs to find the correct deformation minimum with the correct blocking orbital(s).
To this end, one can perform calculations by blocking respectively the orbitals near the Fermi surface and identify the result with the lowest energy as the ground state \cite{Pena-Arteaga2016EPJA,Xia2018ADNDT,Zhang2020PRC}.
This procedure is referred to as ``orbital-fixed blocking'' in the following.
In practice, one can block all orbitals in turn within an energy window around the Fermi energy.
However, in the deformed case, the computational cost of the orbital-fixed blocking calculation is usually high.

One possible blocking procedure to obtain the ground state for an odd-$A$ or odd-odd nucleus is to block the lowest quasiparticle orbital(s) in each iteration, referred to as ``automatic blocking'' in the following.
Since the ground state of an odd-$A$ nucleus in Eq.~\eqref{eq:Phi1} is the one-quasiparticle excitation from the RHB vacuum of an even-even nucleus, a lower quasiparticle energy $E_{k_b}$ of the blocked orbital generally corresponds to a lower total energy.
Obviously the automatic blocking procedure would consume less computational resource than the orbital-fixed blocking procedure.
A similar recipe has been used in the relativistic mean-field calculations with the BCS theory in Ref.~\cite{Geng2004MPLA}.

\begin{figure}[htbp]
	\centering
	\includegraphics[width=0.5\textwidth]{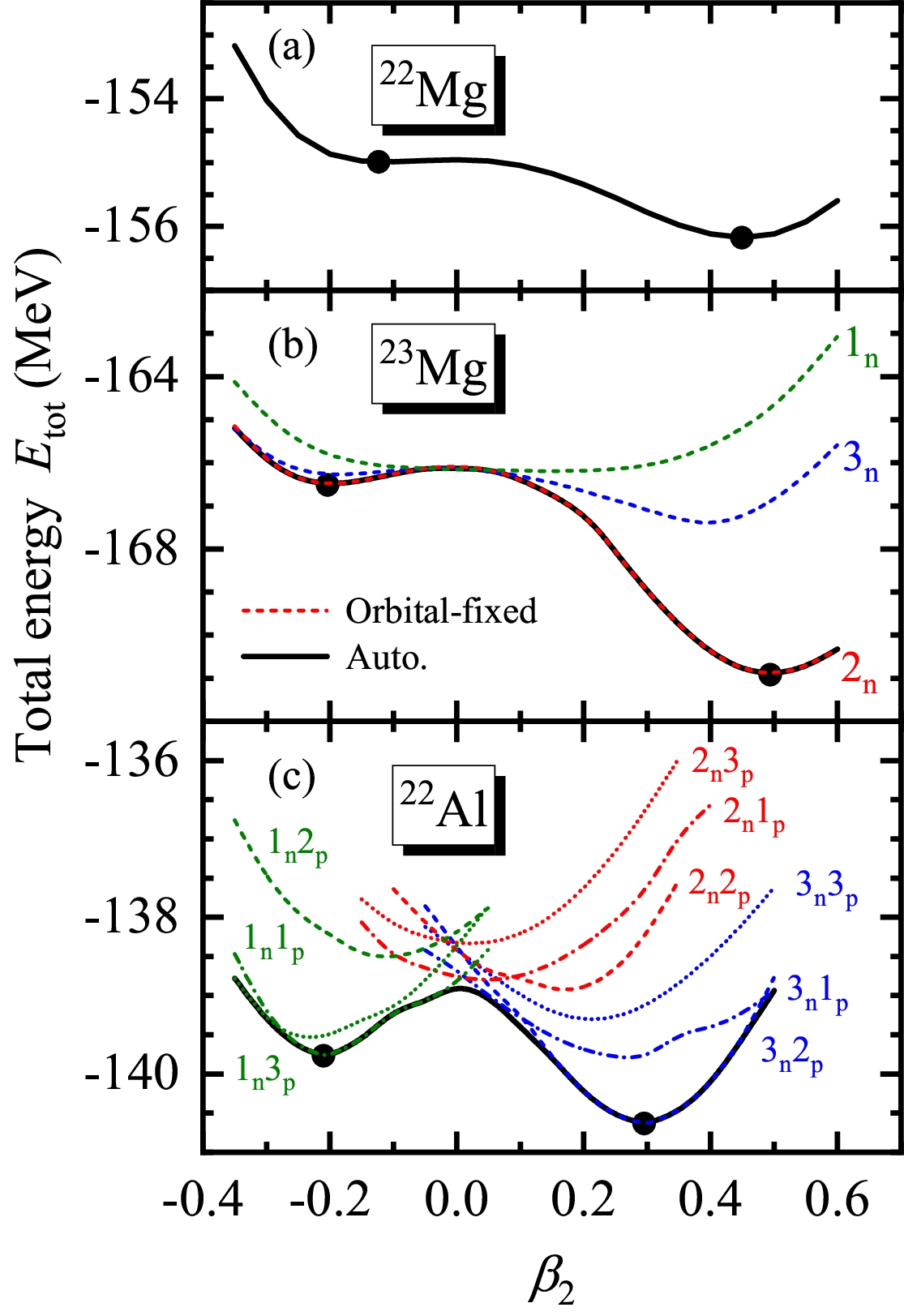}
	\caption{(Color online) Potential energy curves (PECs) of $^{22}$Mg (a), $^{23}$Mg (b) and $^{22}$Al (c) in constrained DRHBc calculations.
	In (b) and (c), the results from the orbital-fixed and automatic blocking procedures are shown with dashed lines and solid lines, respectively.
	The unconstrained minima are also shown with filled circles. }
	\label{fig1}
\end{figure}

\begin{figure}[htbp]
	\centering
	\includegraphics[width=0.5\textwidth]{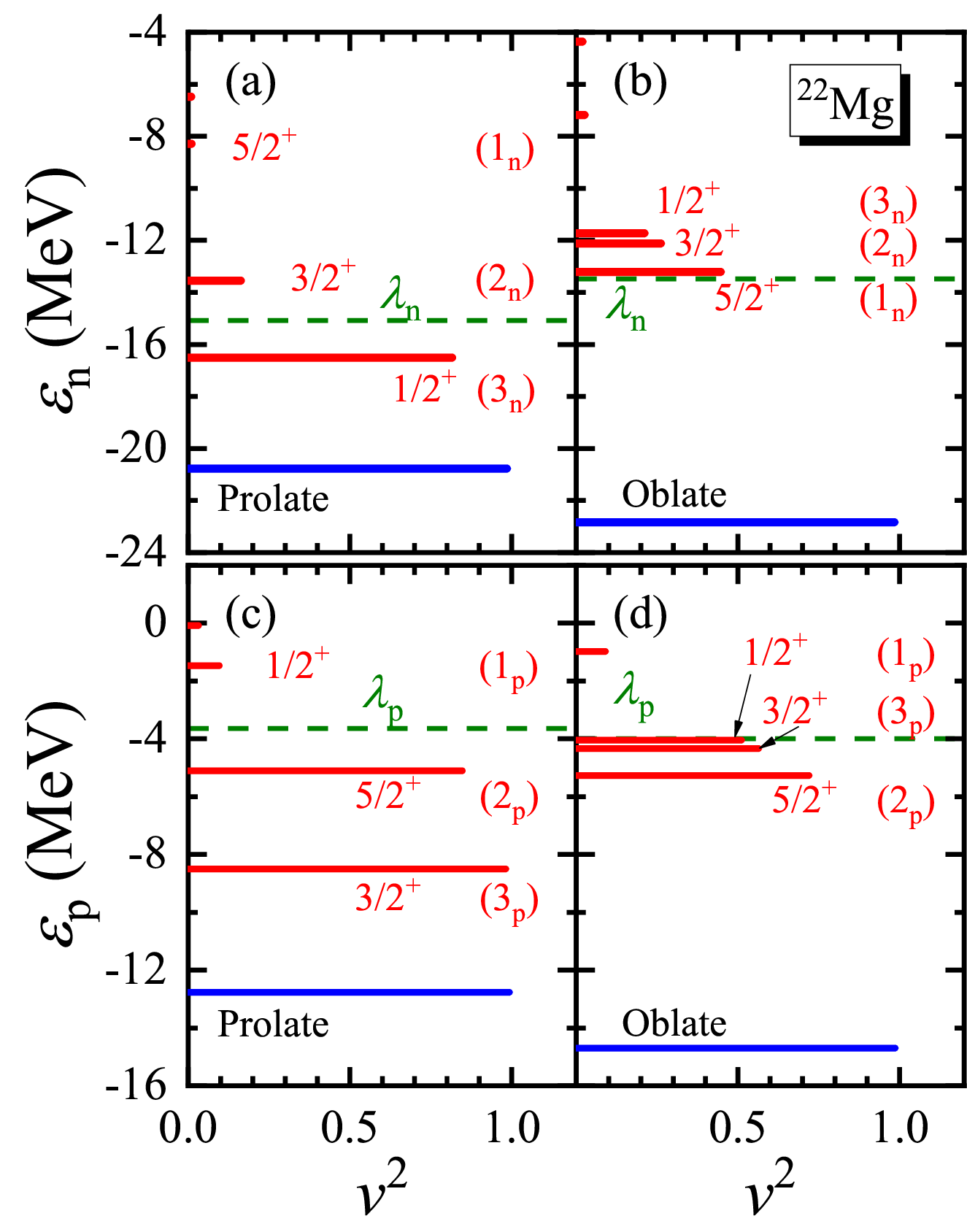}
	\caption{(Color online) Single neutron [(a), (b)] and proton [(c), (d)] orbitals near the Fermi energy in the canonical basis versus the occupation probability $v^2$ for both the prolate [(a), (c)] and oblate [(b), (d)] minima of $^{22}$Mg in the DRHBc calculations.
	The quantum numbers $m^\pi$ of the three orbitals nearest to the Fermi energy are given, and labeled 1, 2 and 3, where the subscript $n$ refers to neutron, and $p$ refers to proton.
	The Fermi energies are shown with the dashed lines.  }
	\label{fig2}
\end{figure}

Taking the odd-$A$ nucleus $^{23}$Mg and odd-odd nucleus $^{22}$Al as examples, the validity of the automatic blocking procedure is checked by comparing the potential energy curves (PECs) with those obtained from the orbital-fixed blocking calculations.
For the orbital-fixed blocking calculations, the procedures are as followed.
\begin{enumerate}
\item The neighboring even-even nucleus $^{22}$Mg is calculated, and the obtained PEC is shown in Fig.~\ref{fig1}(a).
There are two local minima at prolate and oblate sides, respectively.

\item The neutron and proton single-particle orbitals in the canonical basis are shown in Fig.~\ref{fig2} for $^{22}$Mg at both prolate and oblate minima.
Three neutron orbitals near the Fermi energy are labeled as $1_{n}$, $2_{n}$ and $3_{n}$, respectively, and similarly for the proton orbitals.
To identify the ground state for $^{23}$Mg, the orbitals $1_{n}$, $2_{n}$ and $3_{n}$ need to be blocked respectively.
For $^{22}$Al, both the neutron and proton orbitals need to be blocked.

\item For given blocked orbitals, the corresponding PECs are calculated and shown in Figs.~\ref{fig1}(b) and (c).
\end{enumerate}

For $^{23}$Mg, as shown in Fig.~\ref{fig1}(b), the PEC by blocking the orbital $2_n$ is always the lowest, and there are two minima with prolate and oblate deformations, respectively.
The PEC by automatic blocking procedure is always the same as that by blocking the orbital $2_n$.
For $^{22}$Al as shown in Fig.~\ref{fig1}(c), the PEC by blocking the orbitals $3_n$ and $2_p$ is the lowest at the prolate side, and the PEC by blocking the orbitals $1_n$ and $3_p$ is the lowest at the oblate side.
The PEC by automatic blocking procedure coincides with that by blocking the orbitals $3_n$ and $2_p$ at the prolate side, and with that by blocking the orbitals $1_n$ and $3_p$ at the oblate side.

Therefore, it is confirmed that the automatic blocking procedure correctly leads to the ground state for the odd-$A$ nucleus $^{23}$Mg and odd-odd nucleus $^{22}$Al, but significantly reduces the computational cost.

\begin{figure}[htbp]
	\centering
	\includegraphics[width=0.55\textwidth]{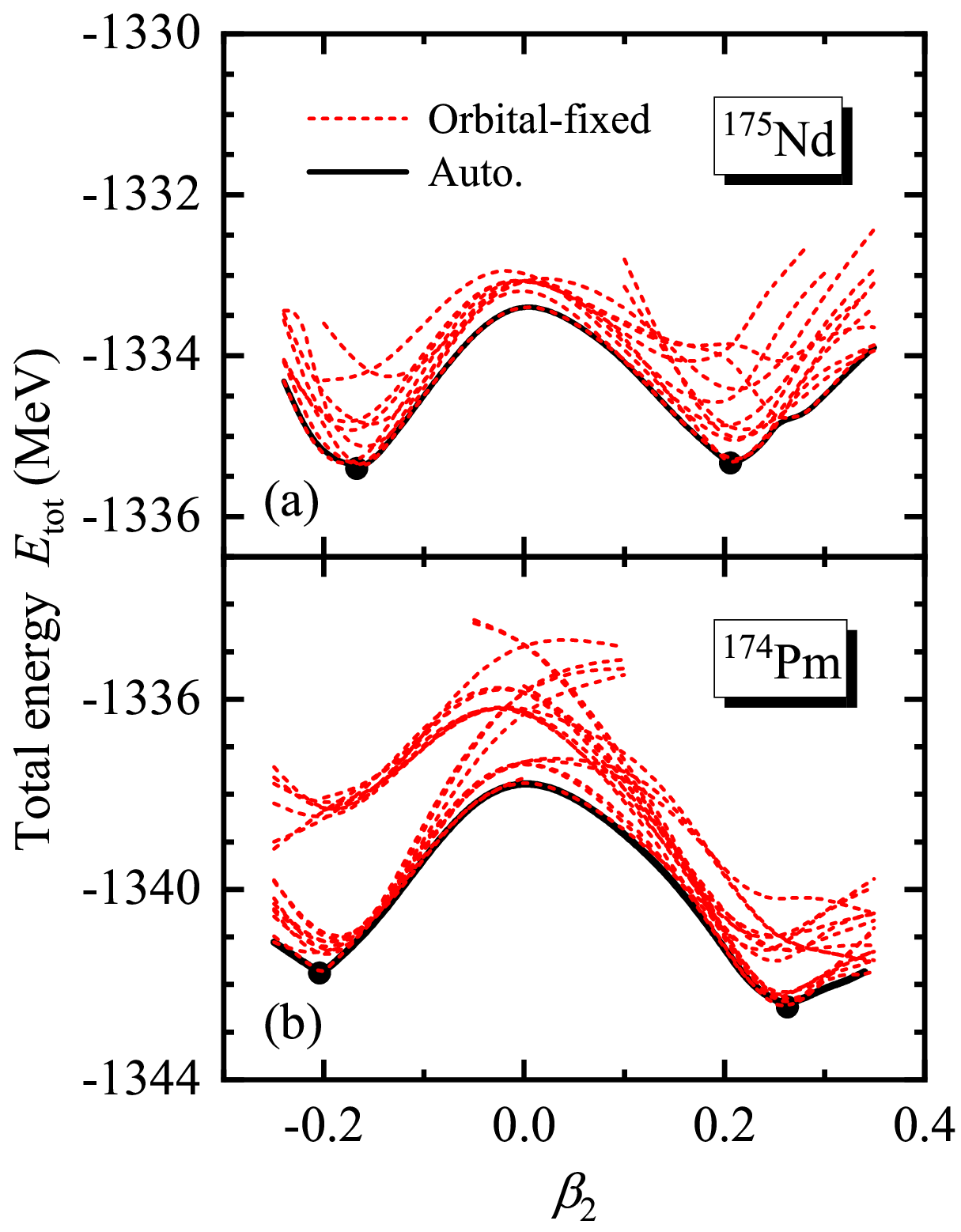}
	\caption{(Color online) Potential energy curves (PECs) of $^{175}$Nd (a) and $^{174}$Pm (b) in constrained DRHBc calculations.
	The results of both the orbital-fixed and automatic blocking procedures are shown with dashed lines and solid lines, respectively.
	The unconstrained minima are also shown with filled circles. }
	\label{fig3}
\end{figure}

The same procedures are applied for the heavy odd-$A$ nucleus $^{175}$Nd and odd-odd nucleus $^{174}$Pm.
With a 2 MeV window around the Fermi energy, the single neutron and proton orbitals for $^{174}$Nd at both prolate and oblate sides are taken as the blocking orbital candidates.
For $^{175}$Nd, the PEC by automatic blocking procedure and 11 lowest PECs by orbital-fixed blocking procedure are shown in Fig.~\ref{fig3}(a).
For $^{174}$Pm, the PEC by automatic blocking procedure and 20 lowest PECs by orbital-fixed blocking procedure are shown in Fig.~\ref{fig3}(b).
It is found the PEC by automatic blocking procedure coincides with the lowest result by orbital-fixed blocking procedure for the odd-$A$ nucleus $^{175}$Nd and odd-odd nucleus $^{174}$Pm.

The automatic blocking procedure is an efficient approach to look for the ground states for odd-$A$ and odd-odd nuclei.
However, if the lowest quasiparticle orbitals are near degenerate, the iteration of the calculation may have difficulty in converging.
In this case, the orbital-fixed blocking procedure is necessary.

\subsection{Convergence check}
\label{subsec:numconv}

The numerical conditions for even-even nuclei have been suggested for DRHBc mass table calculations in Ref.~\cite{Zhang2020PRC}, where the energy cutoff for Woods-Saxon basis is $E_{\mathrm{cut}} = 300$ MeV, the angular momentum cutoff is $J_{\max} = 23/2 ~ \hbar $, and the Legendre expansion truncation for Eq.~\eqref{eq:Legexp} is $\lambda_{\max}=6$ for $Z \leq 80$ and $\lambda_{\max}=8$ for $Z > 80$.
The illustrative nuclei in Ref.~\cite{Zhang2020PRC} include doubly-magic nuclei $^{40}$Ca, $^{100}$Sn and $^{208}$Pb, as well as open-shell nuclei $^{20}$Ne, $^{112}$Mo and $^{300}$Th.
In the following the numerical convergence for odd-$A$ and odd-odd nuclei are checked by taking $^{301}$Th as an example.

\begin{figure}[htbp]
	\centering
	\includegraphics[width=1\textwidth]{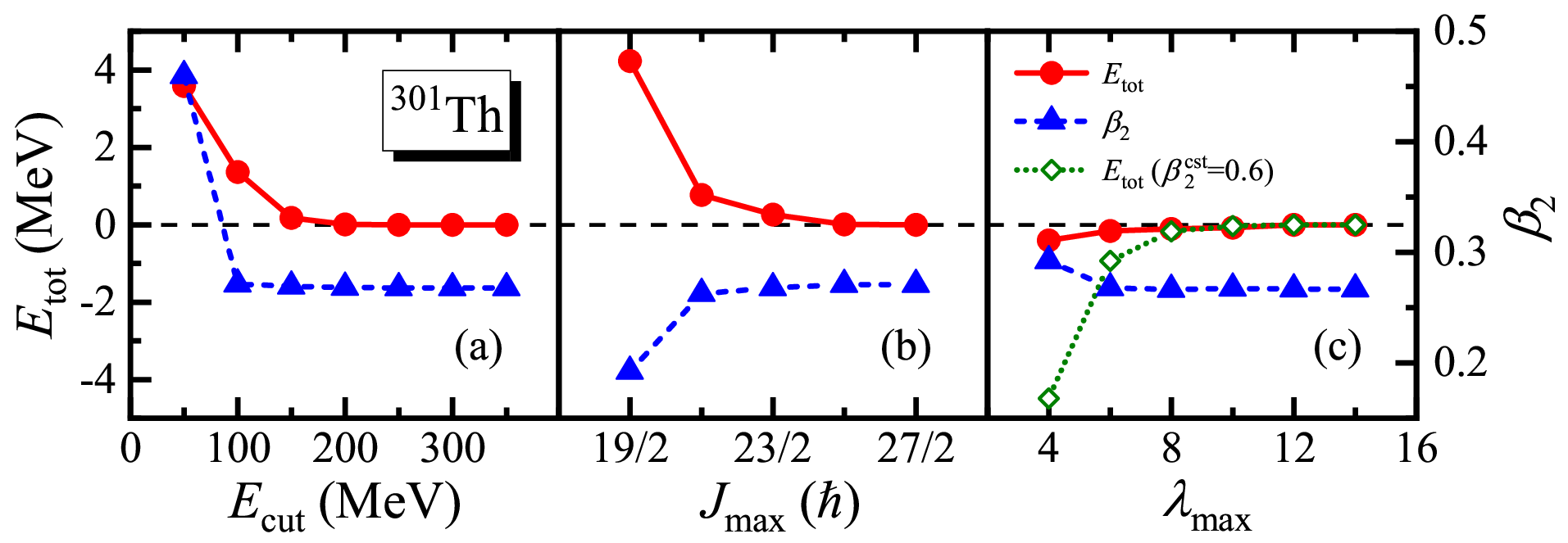}
	\caption{(Color online) Total energy $E_{\mathrm{tot}}$ and quadrupole deformation parameter $\beta_2$ as functions of the energy cutoff $E_{\mathrm{cut}}$ (a), angular momentum cutoff $J_{\max}$ (b) and Legendre expansion truncation $\lambda_{\max}$ (c) for $^{301}$Th from the DRHBc calculations.
	The converged values of the total energy are shifted to zero.
	In panel (c), the result from constrained DRHBc calculations at the quadrupole deformation $\beta_2^{\mathrm{cst}}=0.6$ is also shown to confirm the convergence of $\lambda_{\max}$ at a large deformation.
	Here the pairing correlation is neglected.
	}
	\label{fig4}
\end{figure}

Figure \ref{fig4} shows the changes of total energy $E_{\mathrm{tot}}$ and quadrupole deformation parameter $\beta_2$ with the energy cutoff $E_{\mathrm{cut}}$, angular momentum cutoff $J_{\max}$ and Legendre expansion truncation $\lambda_{\max}$ for $^{301}$Th.
Similar to even-even nuclei \cite{Zhang2020PRC}, the pairing correlation is neglected here in order to avoid renormalizing the strength of the zero-range pairing force to the corresponding model space.
By changing $E_{\mathrm{cut}}$ from 300 to 350 MeV, $\beta_2$ varies by less than 0.0001 and $E_{\mathrm{tot}}$ varies by 0.0007 MeV, showing excellent convergence.
By changing $J_{\max}$ from 23/2 to 27/2 $\hbar$, $\beta_2$ varies by 0.003 and $E_{\mathrm{tot}}$ varies by 0.01\%, i.e., 0.2583 MeV.
By changing $\lambda_{\max}$ from 8 to 14, $E_{\mathrm{tot}}$ varies by 0.1076 MeV and $\beta_2$ varies by 0.0001.
In order to confirm the convergence of $\lambda_{\max}$ at a large deformation, the convergence for $\lambda_{\max}$ is checked in a deformation constrained calculation with $\beta_2^{\mathrm{cst}}=0.6$.
By changing $\lambda_{\max}$ from 8 to 14, $E_{\mathrm{tot}}$ varies by 0.01\%, i.e., 0.1777 MeV.

Therefore, it is concluded that the numerical details in Ref.~\cite{Zhang2020PRC} are still valid in the calculations for odd-$A$ and odd-odd nuclei.
In the following calculations, the numerical details are as follows.
The relativistic density functional PC-PK1 \cite{Zhao2010PRC}, which has turned out to be one of the best density functionals for describing nuclear properties \cite{Zhao2012PRCmass,Lu2015PRC,Agbemava2015PRC}, is adopted.
The box size and mesh size are $R_{\mathrm{box}} = 20$ fm and $\Delta r = 0.1$ fm;
the energy cutoff for the levels in the Fermi sea is $E_{\mathrm{cut}}=300$ MeV, and the number of states in the Dirac sea is taken to be the same as that in the Fermi sea;
the angular momentum cutoff is $J_{\max} = 23/2 ~ \hbar$;
the Legendre expansion truncations in Eq.~\eqref{eq:Legexp} are chosen as $\lambda_{\max} = 6$ and 8 for nuclei with $8 \leq Z \leq 70$ and $72 \leq Z \leq 100$ respectively, and for superheavy nuclei with $102 \leq Z \leq 120$, $\lambda_{\max} = 10$ is adopted \cite{Pan2019IJMPE,Zhang2020PRC};
the pairing strength $V_0 = -325.0 ~ \mathrm{MeV ~ fm}^3$ and the sharp pairing window of 100 MeV are used, which reproduce well the odd-even mass differences for calcium and lead isotopes \cite{Zhang2020PRC}.

\section{Results and discussion}
\label{sec:Nd}

Taking Nd isotopic chain as an example, the ground-state properties, including the binding energy, two-neutron and one-neutron separation energies, Fermi energies, quadrupole deformations, rms radii, and density distributions obtained from the DRHBc calculations are shown in this Section.
The calculated results of even-even Nd isotopes are the same with those in Ref.~\cite{Zhang2020PRC}.
In the calculations for odd-$A$ Nd isotopes, the automatic blocking procedure is adopted and the numerical details are given in Section \ref{subsec:numconv}.
The ground-state properties of Nd isotopes are tabulated in the Appendix \ref{sec:tab}.
The mass number $A$, neutron number $N$, binding energy $E_{\mathrm{B}}$, two-neutron and one-neutron separation energies $S_{2n}$ and $S_n$, rotational correction energy $E_{\mathrm{rot}}$, rms radii $R_n$, $R_p$, $R_m$ and $R_{\mathrm{ch}}$, quadrupole deformations $\beta_{2,n}$, $\beta_{2,p}$ and $\beta_{2}$, Fermi energies $\lambda_n$ and $\lambda_p$, and the quantum numbers $m^\pi$ of the blocked orbital are listed.

\subsection{Binding energy}

\begin{figure}[htbp]
  \centering
  \includegraphics[width=0.6\textwidth]{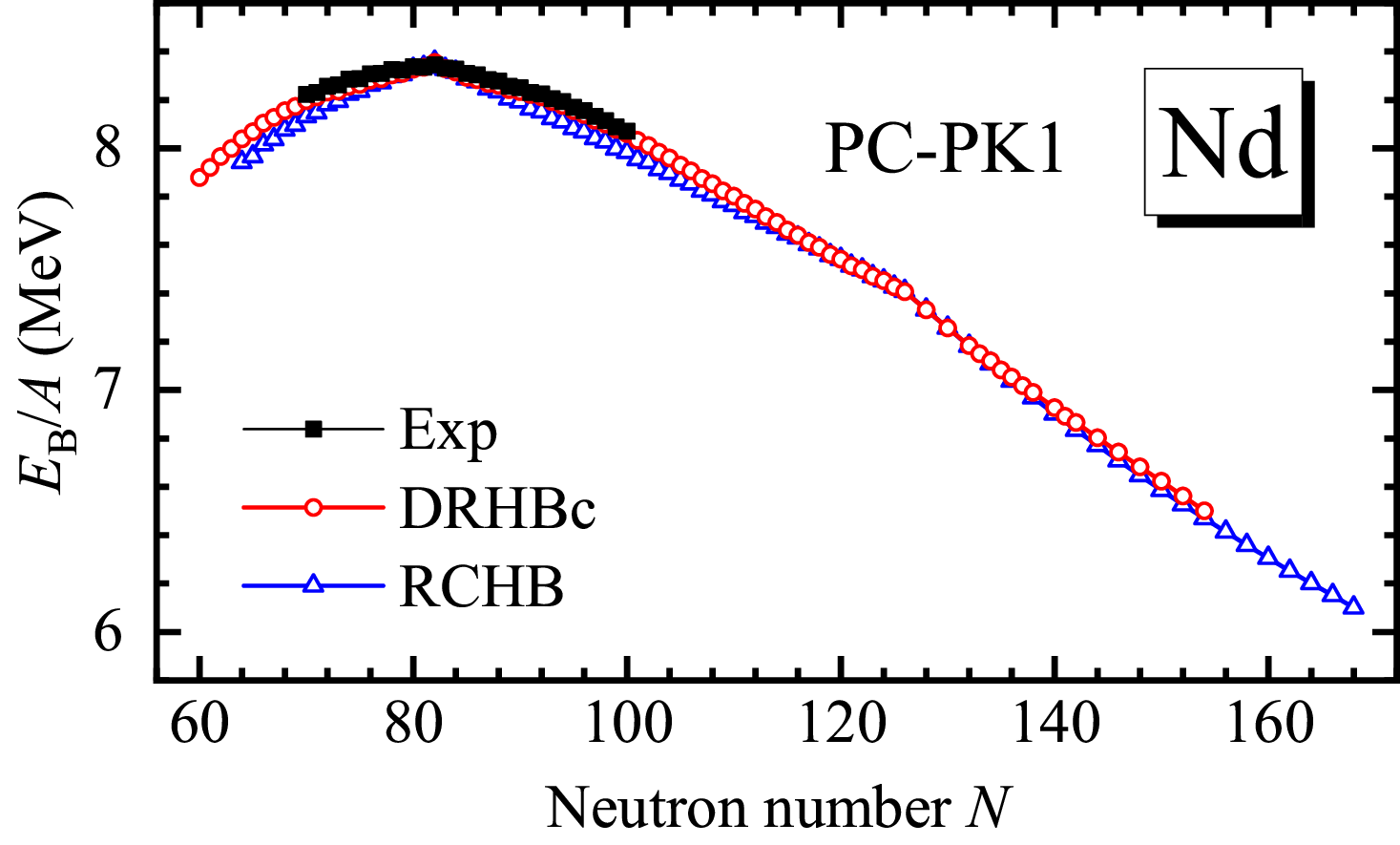}
  \caption{(Color online) Binding energy per nucleon of Nd isotopes from the DRHBc calculations as a function of the neutron number.
  The results in the RCHB mass table \cite{Xia2018ADNDT} and the experimental data from Ref. \cite{Wang2021CPC} are shown for comparison. }
  \label{fig5}
\end{figure}

In Fig. \ref{fig5}, the binding energies per nucleon $E_{\mathrm{B}}/A$ versus the neutron number for Nd isotopes are shown, together with the RCHB results \cite{Xia2018ADNDT} and the available experimental data.
By increasing the neutron number from the proton drip line to $N=82$, $E_{\mathrm{B}}/A$ increases gradually, and from $N=82$ to the neutron drip line, $E_{\mathrm{B}}/A$ decreases gradually.
For both even-even and odd-$A$ isotopes, the data of $E_{\mathrm{B}}/A$ are well reproduced by DRHBc.

\begin{figure}[htbp]
  \centering
  \includegraphics[width=0.6\textwidth]{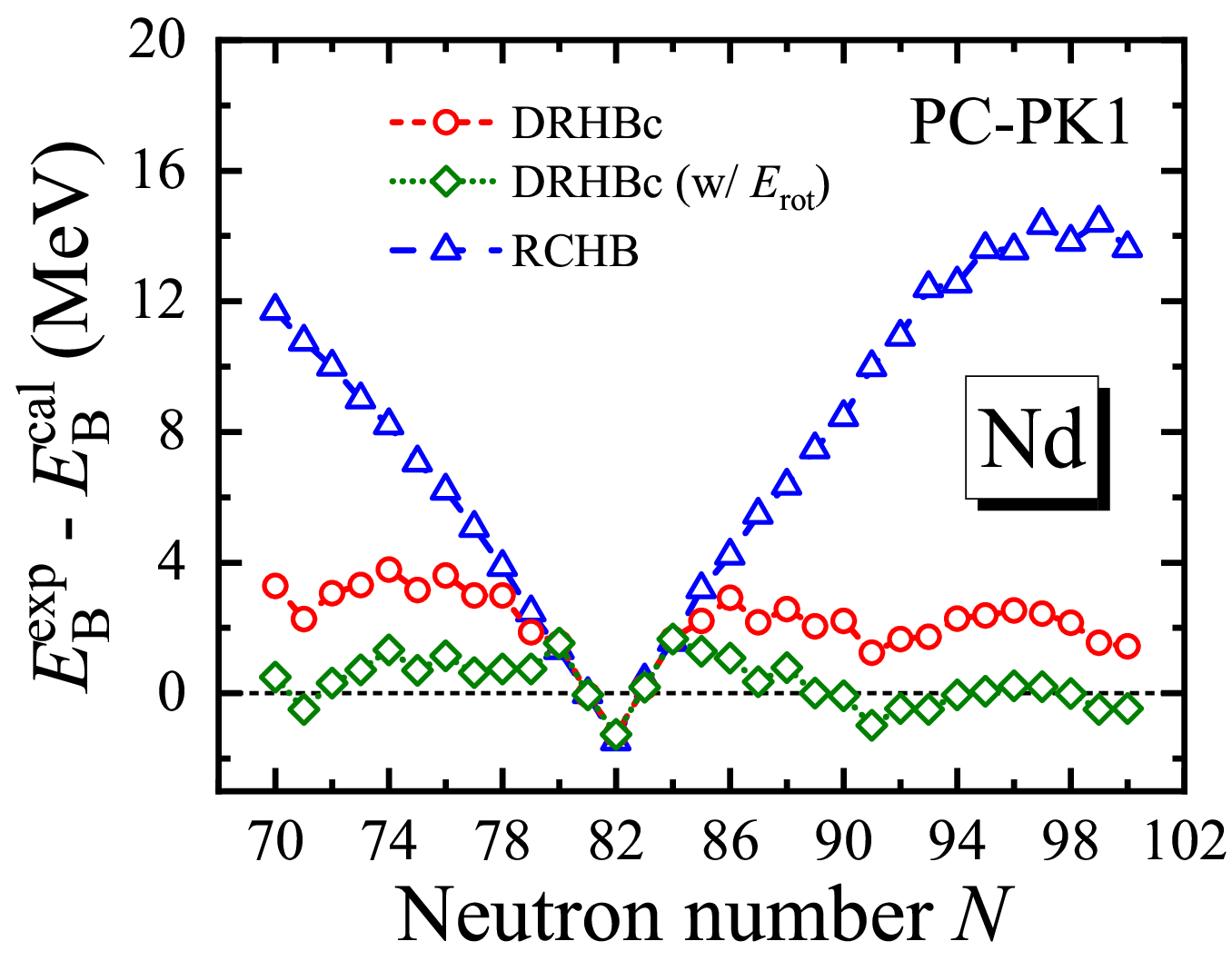}
  \caption{(Color online) The difference between the experimental binding energy \cite{Wang2021CPC} and the DRHBc calculations for Nd isotopes versus the neutron number.
  The results of the DRHBc calculations including the rotational correction energy $E_{\mathrm{rot}}$ and the RCHB mass table \cite{Xia2018ADNDT} are also shown for comparison.
  }
  \label{fig6}
\end{figure}

For a quantitative comparison, Fig.~\ref{fig6} shows the difference between the calculated binding energies and the experimental data.
The rms deviations for the binding energy given by the DRHBc calculations including the rotational correction energy $E_{\mathrm{rot}}$ \cite{Zhao2010PRC,Zhang2020PRC} are 0.90 MeV for even-even isotopes and 0.64 MeV for odd-$A$ isotopes, and the overall rms deviation is 0.78 MeV.
Without $E_{\mathrm{rot}}$, the overall rms deviation is 2.38 MeV, with 2.55 MeV for even-even isotopes and 2.18 MeV for odd-$A$ isotopes.
Without the deformation effect, i.e., in the RCHB results, the overall rms deviation is 9.08 MeV, with 9.12 MeV for even-even isotopes and 9.04 MeV for odd-$A$ isotopes.
These results indicate that including the deformation effect significantly improves the description on nuclear masses for both even-even and odd-$A$ isotopes, and the rotational correction energy can provide further improvements.

\subsection{Neutron separation energies}

\begin{figure}[htbp]
  \centering
  \includegraphics[width=0.7\textwidth]{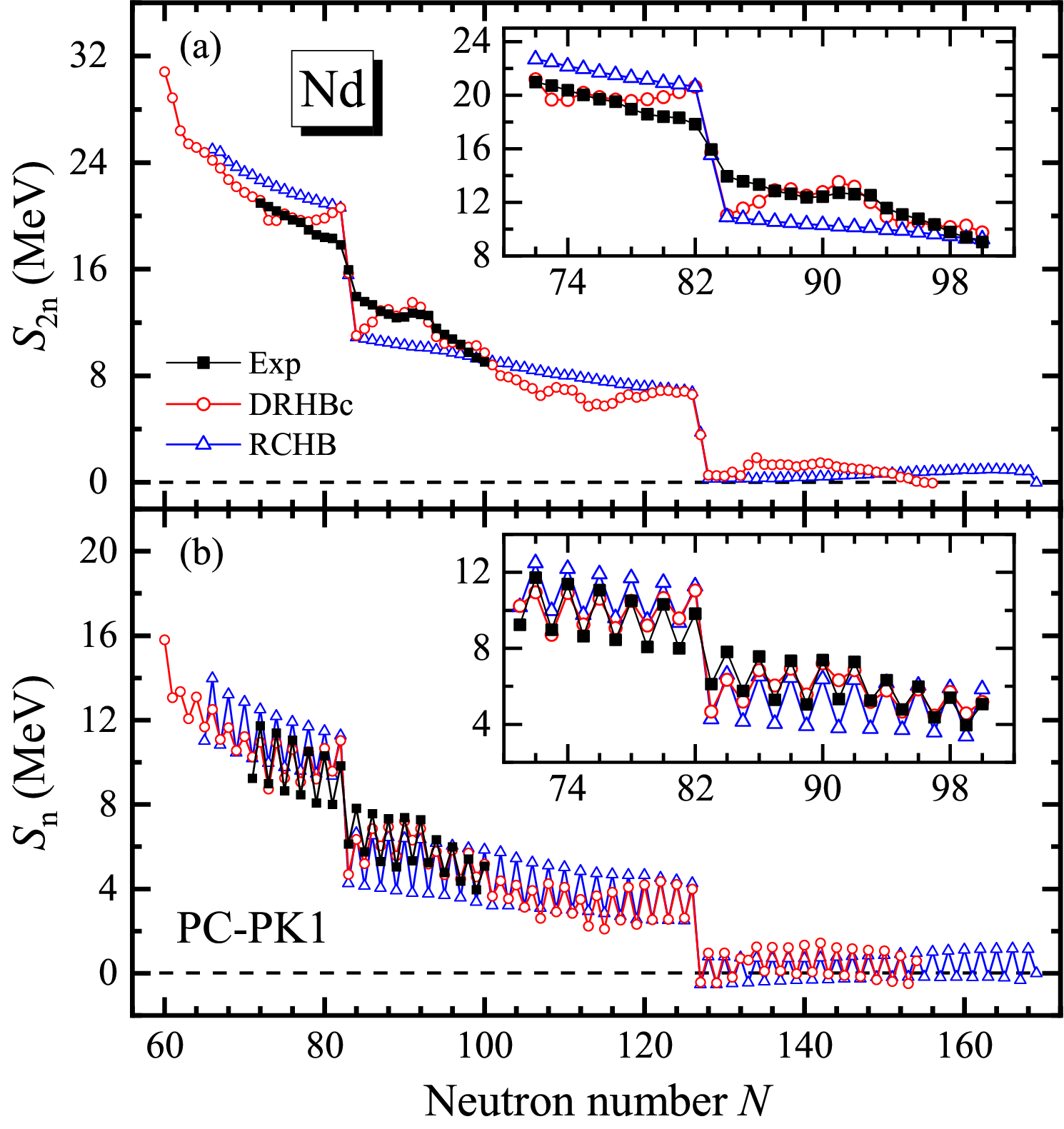}
  \caption{(Color online) Two-neutron (a) and one-neutron (b) separation energies as functions of the
  neutron number for Nd isotopes in the DRHBc calculations.
  The RCHB results \cite{Xia2018ADNDT} and the available data \cite{Wang2021CPC} are also shown for comparison.
  The inset gives the detailed comparison between the calculated results and the data. }
  \label{fig:Nd_Sn}
\end{figure}

From the binding energies, the two-neutron and one-neutron separation energies $S_{2n}$ and $S_{n}$ can be calculated as
\begin{align}
	S_{2n}(Z,N) & = E_{\mathrm{B}}(Z,N) - E_{\mathrm{B}}(Z,N-2), \\
	S_{n}(Z,N)  & = E_{\mathrm{B}}(Z,N) - E_{\mathrm{B}}(Z,N-1).
\end{align}
Figure \ref{fig:Nd_Sn} shows the $S_{2n}$ and $S_{n}$ obtained from the DRHBc calculations, in comparison with the RCHB results and the available data.
By increasing $N$, the $S_{2n}$ generally decreases with smooth transition between the even-even and odd-$A$ isotopes in most cases.
In contrast, the significant odd-even staggering in $S_{n}$ is noticed, i.e., the $S_n$ of an odd-$A$ isotope is lower than the neighboring even-even ones, reflecting the blocking effect of the unpaired odd neutron.
The inset gives a more detailed comparison between the calculated results and the available data.
The rms deviations of $S_{2n}$ and $S_{n}$ from the data are 1.10 and 0.74 MeV for the DRHBc results, which are smaller than 2.04 and 1.10 MeV for the RCHB results, respectively.
This indicates the importance of deformation for the description of two-neutron and one-neutron separation energies.

With the increase of neutron number, when $S_{2n}$ changes from positive to negative, the two-neutron drip line is reached, and the last nucleus with positive $S_{2n}$ is regarded as the two-neutron drip-line nucleus.
The one-neutron drip-line nucleus can be determined according to $S_{n}$ in the same way.
From Fig.~\ref{fig:Nd_Sn}, the two-neutron drip-line nucleus is $^{214}$Nd and the $S_{2n}$ decreases to negative at $N=156$, which has been discussed in Ref.~\cite{Zhang2020PRC}.
The $S_{n}$ decreases to negative at $N=127$, and thus the one-neutron drip-line nucleus is $^{186}$Nd, which has 28 neutrons less than the two-neutron drip-line one,
which reflects the impact of the pairing correlation and blocking effect in odd-$A$ nuclei.
Similar conclusion has also been presented in other works, for example, the two-neutron and one-neutron drip-line nuclei of Nd in the RCHB predictions are $^{228}$Nd and $^{186}$Nd \cite{Xia2018ADNDT}, respectively.

It is found in Fig.~\ref{fig:Nd_Sn} that among the odd-$A$ isotopes beyond the one-neutron drip line $N=126$, the $S_{n}$ becomes positive again at $N=133$, 135, 137 and 141, where the two-neutron separation energy $S_{2n}$ and multi-neutron separation energies $S_{xn}$ ($x=3,4,\dots$) are all positive.
This means that $^{193,195,197,201}$Nd are stable against neutron emissions.
According to the systematic DRHBc calculations, some bound even-even nuclei beyond the two-neutron drip line have been predicted in the regions of $50 \leq Z \leq 70$, $80 \leq Z \leq 90$, and $100 \leq Z \leq 120$, and the underlying mechanisms have been investigated \cite{Zhang2021PRC,Pan2021PRC,He2021CPC,Zhang2022ADNDT}.
Here, a similar phenomenon is predicted in the Nd isotopic chain, but for the odd-$A$ nuclei beyond the one-neutron drip line.

\subsection{Fermi energy}

\begin{figure}[htbp]
  \centering
  \includegraphics[width=0.65\textwidth]{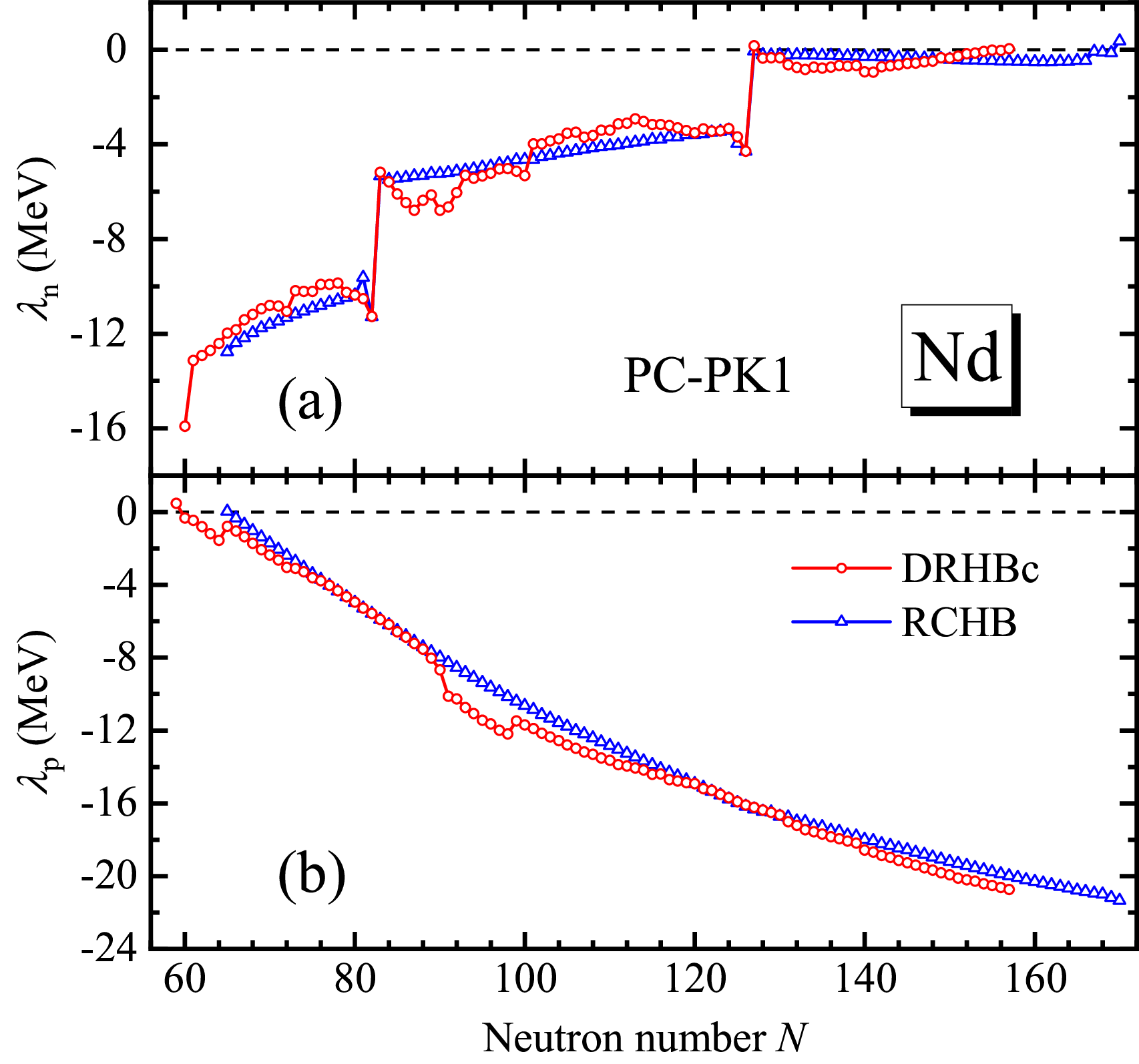}
  \caption{(Color online) Neutron (a) and proton (b) Fermi energies for Nd isotopes in the DRHBc calculations versus the neutron number.
  The results from the RCHB mass table \cite{Xia2018ADNDT} are shown for comparison. }
  \label{fig8}
\end{figure}

The Fermi energy represents, in the mean-field level, the change of the total energy against the particle number \cite{Ring1980NMBP} and also carries information about the nucleon drip line.
A negative Fermi energy usually corresponds to a positive separation energy of a bound nucleus.
Figure \ref{fig8} shows the neutron and proton Fermi energies from DRHBc, in comparison with the RCHB results \cite{Xia2018ADNDT}.
If the pairing energy vanishes, the Fermi energy is chosen to be the energy of the last occupied single-particle level.
It is found that for most Nd isotopes, the changes of Fermi energies between even-even and odd-$A$ ones are smooth.
In Fig.~\ref{fig8}(a), all the isotopes with $S_{2n}>0$ and $S_n>0$ have negative $\lambda_n$.
In Fig.~\ref{fig8}(b), $\lambda_p$ becomes negative at $N=60$, and decreases with the neutron number in most cases.
According to the proton Fermi energy, the proton drip-line nucleus is $^{120}$Nd.
Due to the deformation effect, the evolutions of the Fermi energies obtained by the DRHBc calculations are distinct from the RCHB results \cite{Zhang2020PRC}.
Specifically, the irregularities observed between shell closures are related to the deformation, which is shown in Section \ref{subsec:defor}.

\subsection{Quadrupole deformation}
\label{subsec:defor}

\begin{figure}[htbp]
  \centering
  \includegraphics[width=0.8\textwidth]{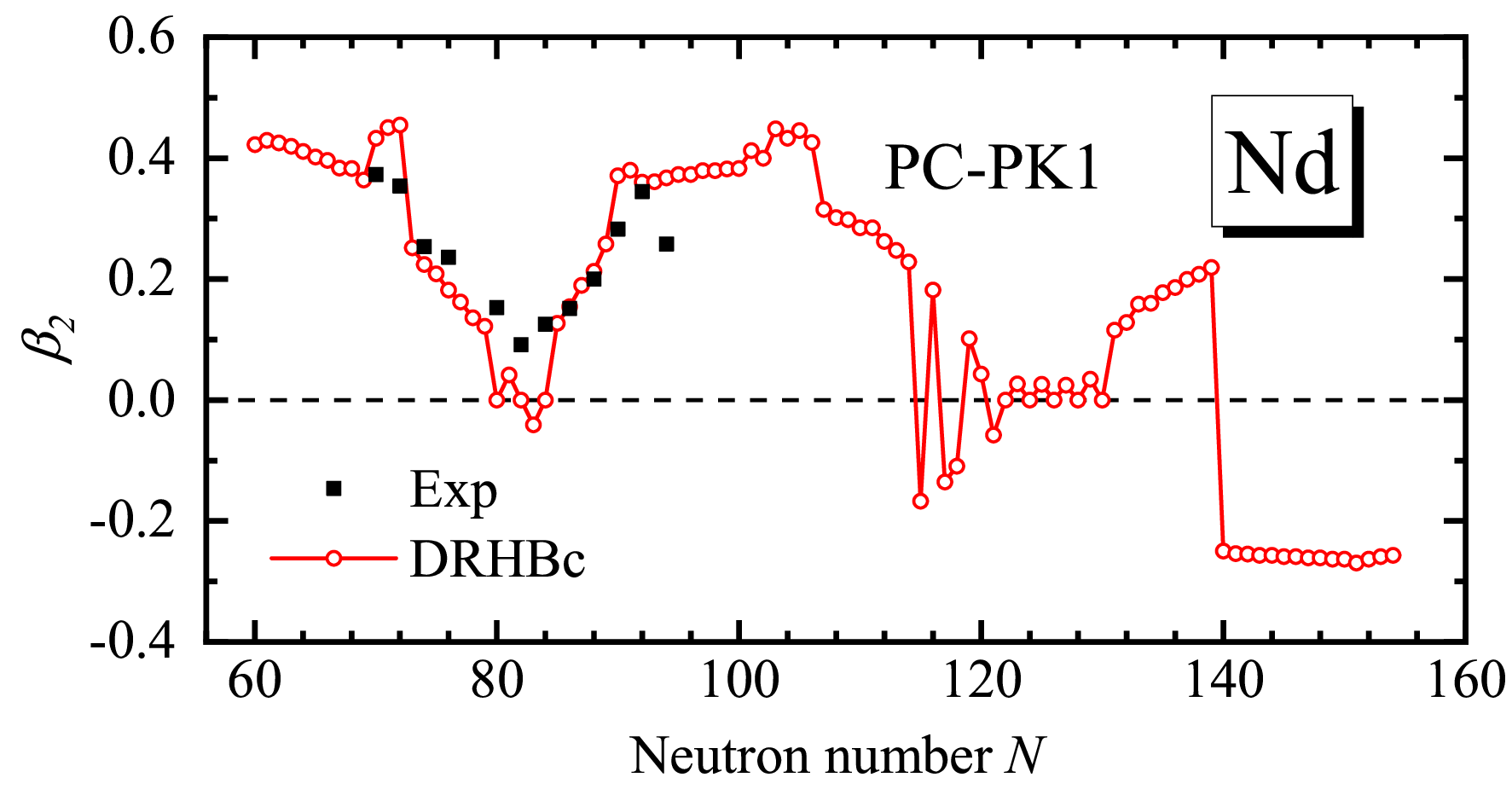}
  \caption{(Color online) Quadrupole deformation as a function of the neutron number in the DRHBc calculations for Nd isotopes.
  The available data \cite{Pritychenko2016ADNDT}, only the absolute values, for even-even nuclei are shown for comparison. }
  \label{fig:Nd_beta}
\end{figure}

The ground-state quadrupole deformation parameters $\beta_2$ of Nd isotopes are shown in Fig.~\ref{fig:Nd_beta} and compared with experimental data \cite{Pritychenko2016ADNDT}.
The data are only available in the absolute values for even-even nuclei and are determined from the observed $B(E2,0^+ \to 2^+)$ by assuming that the nucleus is a rigid rotor.
In Fig.~\ref{fig:Nd_beta}, the calculated results for even-even nuclei agree well with the data, and the neighboring odd-$A$ nuclei also follow the trend.
The evolution of $\beta_2$ between most even-even and odd-$A$ Nd isotopes is smooth.
While it is interesting to note the strong shape oscillations from $N=114$ to 122, and sudden shape changes at $N=118$ and 140.

\begin{figure}[htbp]
	\centering
	\includegraphics[width=0.75\textwidth]{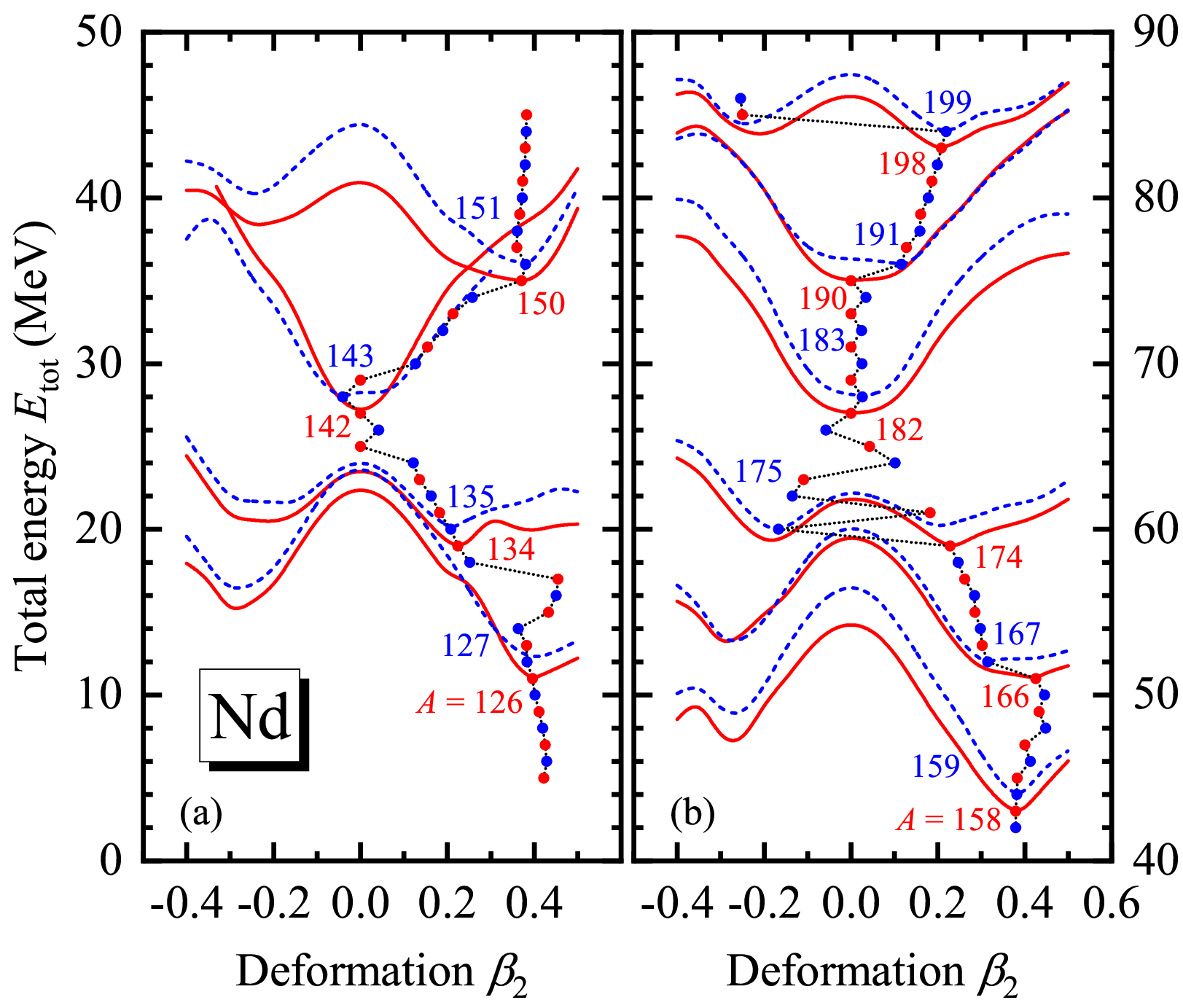}
	\caption{(Color online) Evolution of the potential energy curves of even-even isotopes $^{126,134,142,\dots,198}$Nd (red solid lines) and odd-$A$ isotopes $^{127,135,143,\dots,199}$Nd (blue dashed lines) from constrained DRHBc calculations.
	For clarity reasons, the curves have been scaled to the energy of their ground states and have been shifted upward by 1 MeV per increasing 1 neutron.
	The evolution of the ground-state deformation is shown with the red and blue circles for even-even and odd-$A$ isotopes, respectively.
	}
	\label{fig:Nd_PEC}
\end{figure}

The evolution of deformation can be better understood with the potential energy curve (PEC).
Figure \ref{fig:Nd_PEC} displays the PECs of selected even-even isotopes $^{126,134,142,\dots,198}$Nd and odd-$A$ isotopes $^{127,135,143,\dots,199}$Nd.
The evolution of ground-state deformation with the neutron number is guided by the dotted line.
The global minima of the PECs obtained from constrained calculations are consistent with the ground states from unconstrained ones, guaranteeing the self-consistency of the DRHBc calculations.
It is noted that the PEC of an odd-$A$ Nd isotope is similar to that of its even-even neighbors in most cases, corresponding to their similar ground-state deformations.
The sudden changes of ground-state deformation in Fig.~\ref{fig:Nd_beta} can be understood from the PECs in Fig.~\ref{fig:Nd_PEC}.
For example, both the PECs of $^{174,175}$Nd have two local minima with one at prolate side and another at oblate side, and their energies are close to each other.
The prolate minimum is lower in $^{174}$Nd, whereas the oblate one is lower in $^{175}$Nd, leading to the staggering at $N=114$ in Fig.~\ref{fig:Nd_beta}.
In addition, for the nuclei with two or more minima with similar total energy, possible shape coexistence and/or triaxial deformation is expected \cite{Heyde2011RMP}.

\subsection{Rms radii}

\begin{figure}[htbp]
  \centering
  \includegraphics[width=0.65\textwidth]{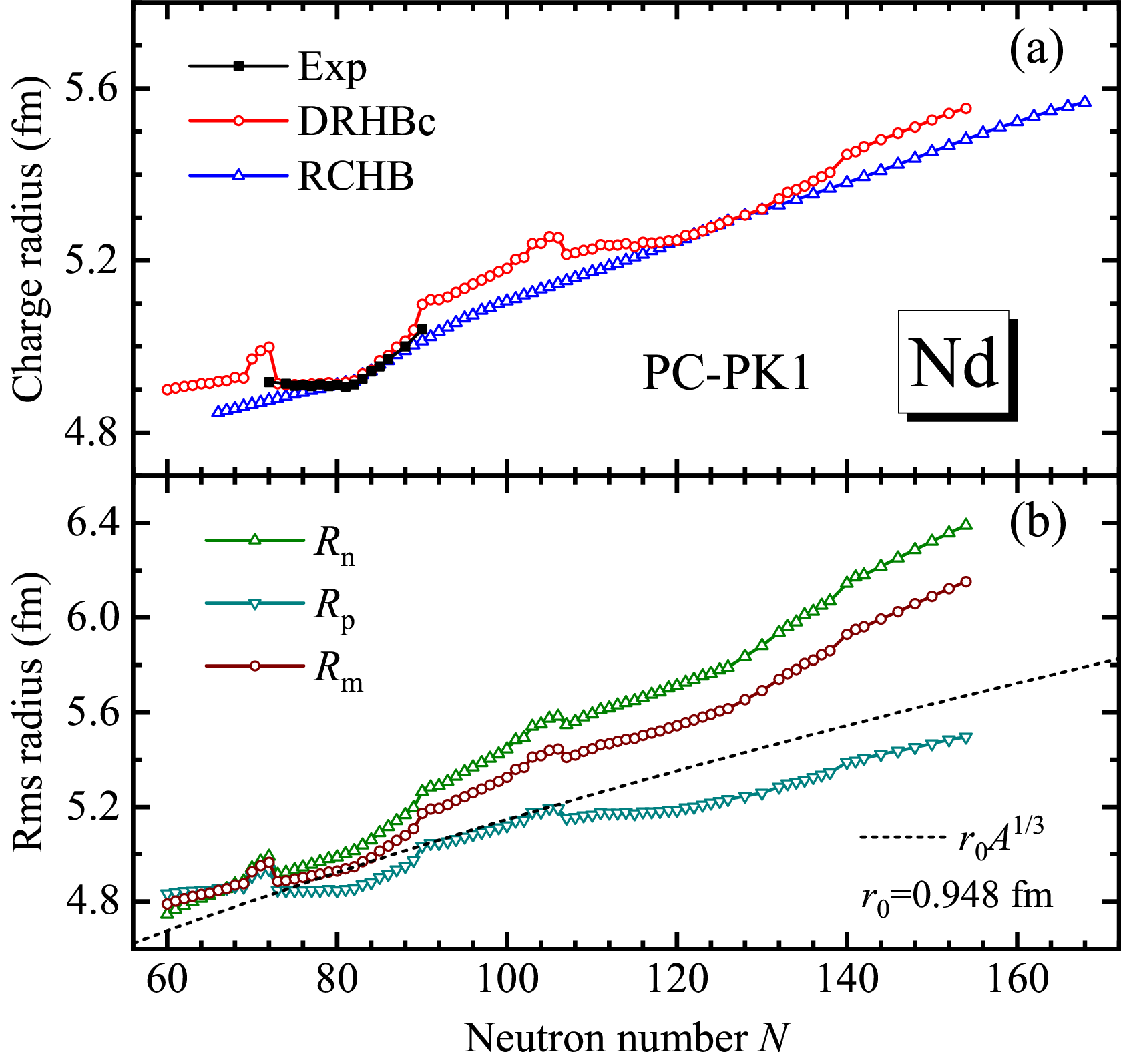}
  \caption{(Color online) (a) Charge radius as a function of the neutron number in the DRHBc calculations for Nd isotopes.
  The results in the RCHB mass table \cite{Xia2018ADNDT} and available data from Ref.~\cite{Angeli2013ADNDT} are shown for comparison.
  (b) Rms neutron radius, proton radius, and matter radius as functions of the neutron number in the DRHBc calculations for Nd isotopes.
  The empirical matter radii $r_0 A^{1/3}$, in which $r_0 = 0.948$ fm determined by $^{142}$Nd, are also shown to guide the eye. }
  \label{fig:Nd_rms}
\end{figure}

Figure \ref{fig:Nd_rms} shows the charge radius as well as the rms neutron radius, proton radius and matter radius as functions of the neutron number in the DRHBc calculations for Nd isotopes.
The RCHB results \cite{Xia2018ADNDT} and available data \cite{Angeli2013ADNDT} for charge radius are also plotted for comparison.
For both even-even and odd-$A$ isotopes, DRHBc can well reproduce the data of charge radius, including the kink at $N=82$ corresponding to the shell closure.
It is also noted that the rms matter radius significantly deviates from the empirical $A^{1/3}$-curve at a large $N$, which may indicate some underlying exotic structures \cite{Zhang2020PRC}.

\subsection{Neutron density distribution}

\begin{figure}[h!]
  \centering
  \includegraphics[width=0.43\textwidth]{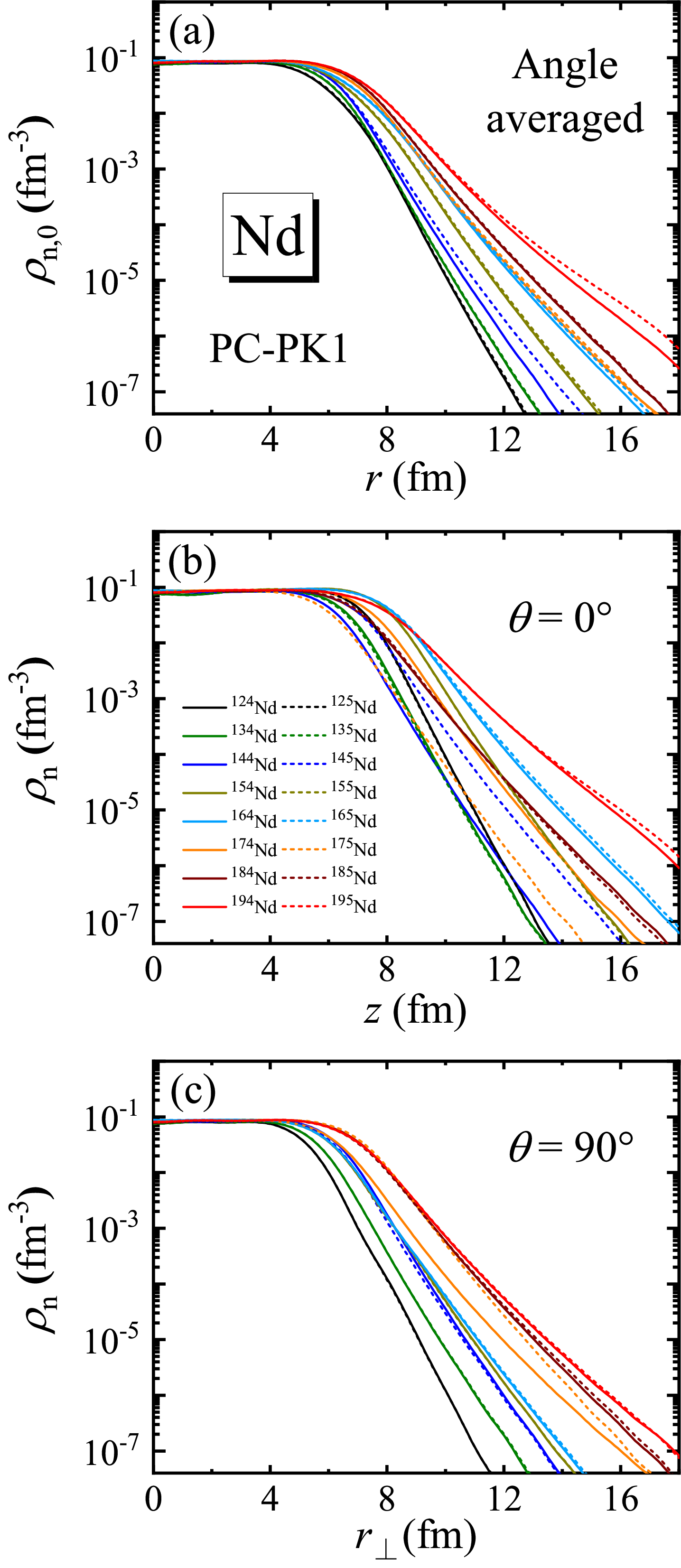}
  \caption{(Color online) (a) Angle-averaged neutron density distribution, i.e., the spherical component, (b) the neutron density distribution along the symmetry axis $z$ ($\theta = 0^\circ$), and (c) the neutron density distribution perpendicular to the symmetry axis with $r_\perp = \sqrt{x^2+y^2}$ ($\theta = 90^\circ$), for selected even-even isotopes $^{124,134,\dots,194}$Nd (solid lines) and odd-$A$ isotopes $^{125,135,\dots,195}$Nd (dashed lines) in the DRHBc calculations. }
  \label{fig12}
\end{figure}

\begin{figure}[h!]
  \centering
  \includegraphics[width=0.6\textwidth]{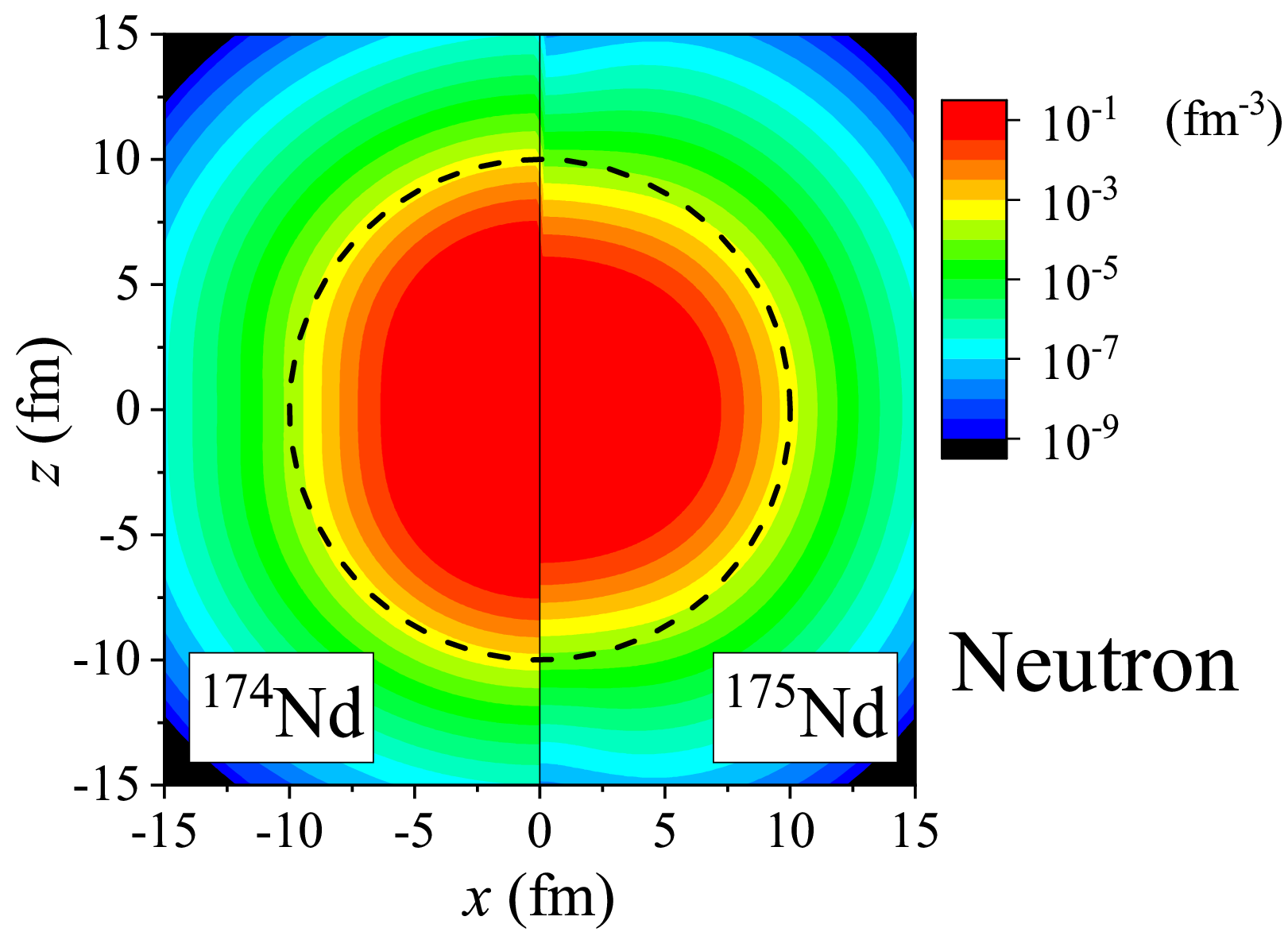} 
  \caption{(Color online) Neutron density distributions for $^{174}$Nd ($x<0$) and $^{175}$Nd ($x>0$) with $z$ axis as the symmetry axis.
  A dashed circle is drawn to guide the eye. }
  \label{fig:Nd_blk_dens2Dneu}
\end{figure}

Figure \ref{fig12} shows the neutron density distributions of selected even-even isotopes $^{124,134,\dots,194}$Nd and odd-$A$ isotopes $^{125,135,\dots,195}$Nd.
The angle-averaged neutron density $\rho_{n,0}$, i.e., the spherical component ($\lambda=0$) in Eq.~\eqref{eq:Legexp}, is shown in Fig.~\ref{fig12}(a);
the neutron density distributions along and perpendicular to the symmetry axis $z$, i.e., $\theta = 0^\circ$ and $\theta = 90^\circ$, are shown in Figs.~\ref{fig12}(b) and (c), respectively.
By increasing $N$, the angle-averaged density in Fig.~\ref{fig12}(a) increases monotonically.
While in Figs.~\ref{fig12}(b) and (c) for $\theta = 0^\circ$ and $90^\circ$, the monotonicity is not obvious, which is related to deformation effect \cite{Zhang2020PRC}.
For example, the densities of $^{184}$Nd and $^{185}$Nd are significantly less than those of $^{164}$Nd and $^{165}$Nd at $\theta=0^\circ$ for $r>6$ fm, because the deformations $\beta_2$ of $^{184}$Nd and $^{185}$Nd are less than 0.03, while those of $^{164}$Nd and $^{165}$Nd are larger than 0.40, as shown in Fig.~\ref{fig:Nd_beta}.
For most odd-$A$ isotopes in Fig.~\ref{fig12} the density distributions are close to those of their even-even neighbors, usually with a slight increase, but there are also some exceptions.
For $^{145}$Nd, the density at $\theta=0^\circ$ extends much farther than that of $^{144}$Nd, which corresponds to the sudden increase of deformation from $\beta_2 = 0$ at $^{144}$Nd to $\beta_2 = 0.13$ at $^{145}$Nd, as shown in Fig.~\ref{fig:Nd_beta}.
Another similar case can be found in $^{174}$Nd and $^{175}$Nd, where $^{174}$Nd is a prolate one with $\beta_2=0.228$ and shows much more significant density at $\theta=0^\circ$ than $^{175}$Nd, while $^{175}$Nd is an oblate one with $\beta_2=-0.167$ and its density at $\theta=90^\circ$ is larger.
In Fig. \ref{fig:Nd_blk_dens2Dneu} the neutron density distributions in the $xz$-plane for $^{174}$Nd and $^{175}$Nd are presented, and the discussion above on density distribution and deformation can be shown in a more clear way.
Furthermore, the angle-averaged density of $^{195}$Nd in Fig.~\ref{fig12}(a) shows an obvious increase in the diffuseness in comparison with $^{194}$Nd, which indicates possible exotic structures of neutron halo or skin \cite{Zhang2020PRC}.

\newpage

\newpage

\newpage

\section{Summary}
\label{sec:summary}

In summary, the DRHBc theory including the blocking effect based on the point-coupling functional is extended to describe the odd-$A$ and odd-odd nuclei.
The blocking procedure and the numerical details are examined.
Taking the neodymium isotopic chain as an example, the ground-state properties are investigated.

The automatic blocking procedure by blocking the lowest quasiparticle orbital is found to be an efficient approach to look for the ground states for odd-$A$ and odd-odd nuclei.
Convergence checks are performed and the numerical details in Ref.~\cite{Zhang2020PRC} are confirmed to be valid in the calculations for odd-$A$ and odd-odd nuclei.

Taking neodymium isotopic chain as an example, DRHBc calculations with the density functional PC-PK1 for both even-even and odd-$A$ isotopes have been performed.
The rms deviation for the binding energy given by the DRHBc calculations including the rotational correction energy is 0.78 MeV.
The importance of the deformation effect on one- and two-neutron separation energies is demonstrated.
The predicted proton drip-line nucleus is $^{120}$Nd.
The predicted one-neutron and two-neutron drip-line nuclei are $^{186}$Nd and $^{214}$Nd, respectively.
The predicted possible bound odd-$A$ nuclei beyond the one-neutron drip line include $^{193,195,197,201}$Nd.
The evolution of quadrupole deformation between most even-even and odd-$A$ Nd isotopes is smooth with the increase of neutron number.
Sudden deformation changes can be understood by the potential energy curve.
The experimental rms charge radii are reproduced well, including the kink at $N=82$ corresponding to the shell closure.
Near the one-neutron drip line, the significant increase in the rms matter radius and in the diffuseness of neutron density distribution indicates possible exotic structures.

\begin{acknowledgments}

This work was partly supported by the National Natural Science Foundation of China (Grants No. 11935003, No. 11875075, No. 11975031, No. 12141501, No. 12047503, No. 11961141004, No. 12070131001, No. U2032138, No. 11775112, No. 12075085, and No. 12147219), the National Key R\&D Program of China (Contracts No. 2017YFE0116700, and No. 2018YFA0404400), the Strategic Priority Research Program of Chinese Academy of Sciences (Grants No. XDB34010000 and No. XDPB15), the China Postdoctoral Science Foundation (Grants No. 2021M700256), High-performance Computing Platform of Peking University, and High-performance Computing Platform of Anhui University. 
YBC and CHL were supported by National Research Foundation of Korea (NRF) grants funded by the Korea government (Ministry of Science and ICT and Ministry of Education)  (No. 2016R1A5A1013277 and No. 2018R1D1A1B07048599) and the National Supercomputing Center with supercomputing resources including technical support (KSC-2021-RND-0076).

\end{acknowledgments}

\appendix
\section{Tabulation of ground-state properties}
\label{sec:tab}

The ground-state properties of Nd isotopes are tabulated in Table \ref{tab:Nd}.

\newpage
\begin{landscape}
\pagestyle{empty}

\begin{longtable}{rrrrrrrrrrrrrrrrrr}
\caption{ Ground-state properties of neodymium isotopes calculated by the DRHBc theory, in comparison with the available data of masses and charge radii. 
In addition, the data labeled with underline mean the nucleus is unbound.
} 
\label{tab:Nd} \\
\hline 
\multirow{2}{*}{$A$} & \multirow{2}{*}{$N$} & $E_{\mathrm{B}}^{\mathrm{Cal}}$ & $E_{\mathrm{B}}^{\mathrm{Exp}}$ & $S_{2n}$ & $S_n$ & $E_{\mathrm{rot}}$ & $R_n$ & $R_p$ & $R_m$ & $R_{\mathrm{ch}}^{\mathrm{Cal}}$ & $R_{\mathrm{ch}}^{\mathrm{Exp}}$ & \multirow{2}{*}{$\beta_{2,n}$} & \multirow{2}{*}{$\beta_{2,p}$} & \multirow{2}{*}{$\beta_2$} & $\lambda_n$ & $\lambda_p$ & \multirow{2}{*}{$m^\pi(N)$} \\
 &  & (MeV) & (MeV) & (MeV) & (MeV) & (MeV) & (fm) & (fm) & (fm) & (fm) & (fm) &  &  &  & (MeV) & (MeV) & \\
\hline \endfirsthead

\hline
\multirow{2}{*}{$A$} & \multirow{2}{*}{$N$} & $E_B^{\mathrm{Cal}}$ & $E_B^{\mathrm{Exp}}$ & $S_{2n}$ & $S_n$ & $E_{\mathrm{rot}}$ & $R_n$ & $R_p$ & $R_m$ & $R_{\mathrm{ch}}^{\mathrm{Cal}}$ & $R_{\mathrm{ch}}^{\mathrm{Exp}}$ & \multirow{2}{*}{$\beta_{2,n}$} & \multirow{2}{*}{$\beta_{2,p}$} & \multirow{2}{*}{$\beta_2$} & $\lambda_n$ & $\lambda_p$ & \multirow{2}{*}{$m^\pi(N)$} \\
 &  & (MeV) & (MeV) & (MeV) & (MeV) & (MeV) & (fm) & (fm) & (fm) & (fm) & (fm) &  &  &  & (MeV) & (MeV) & \\
\hline \endhead

118 & 58 & 914.57 &  &  &  & 2.53 & 4.707 & 4.826 & 4.768 & 4.892 &  & 0.395 & 0.427 & 0.411 & $-$15.62 & \underline{0.75} & $ 0^+ $ \\
119 & 59 & 929.60 &  &  & 15.03 & 2.31 & 4.716 & 4.823 & 4.770 & 4.889 &  & 0.386 & 0.418 & 0.403 & $-$15.66 & \underline{0.49} & $ 3/2^- $ \\
120 & 60 & 945.40 &  & 30.84 & 15.81 & 2.42 & 4.745 & 4.833 & 4.789 & 4.899 &  & 0.411 & 0.433 & 0.422 & $-$15.92 & $-$0.32 & $ 0^+ $ \\
121 & 61 & 958.46 &  & 28.86 & 13.06 & 2.59 & 4.767 & 4.838 & 4.802 & 4.903 &  & 0.422 & 0.436 & 0.429 & $-$13.13 & $-$0.47 & $ 5/2^- $ \\
122 & 62 & 971.81 &  & 26.41 & 13.35 & 2.66 & 4.784 & 4.842 & 4.812 & 4.907 &  & 0.418 & 0.431 & 0.425 & $-$12.93 & $-$0.80 & $ 0^+ $ \\
123 & 63 & 983.87 &  & 25.41 & 12.06 & 2.50 & 4.799 & 4.844 & 4.821 & 4.910 &  & 0.414 & 0.426 & 0.420 & $-$12.71 & $-$1.20 & $ 5/2^+ $ \\
124 & 64 & 996.96 &  & 25.15 & 13.09 & 2.51 & 4.813 & 4.848 & 4.830 & 4.913 &  & 0.404 & 0.418 & 0.411 & $-$12.41 & $-$1.55 & $ 0^+ $ \\
125 & 65 & 1008.63 &  & 24.76 & 11.67 & 2.25 & 4.825 & 4.848 & 4.836 & 4.914 &  & 0.394 & 0.409 & 0.402 & $-$11.98 & $-$0.78 & $ 3/2^+ $ \\
126 & 66 & 1021.14 &  & 24.18 & 12.51 & 2.40 & 4.841 & 4.854 & 4.847 & 4.919 &  & 0.389 & 0.405 & 0.396 & $-$11.84 & $-$1.05 & $ 0^+ $ \\
127 & 67 & 1032.23 &  & 23.60 & 11.09 & 2.25 & 4.854 & 4.855 & 4.854 & 4.920 &  & 0.376 & 0.393 & 0.384 & $-$11.41 & $-$1.36 & $ 1/2^+ $ \\
128 & 68 & 1043.88 &  & 22.74 & 11.65 & 2.57 & 4.874 & 4.863 & 4.869 & 4.928 &  & 0.375 & 0.391 & 0.383 & $-$11.18 & $-$1.72 & $ 0^+ $ \\
129 & 69 & 1054.43 &  & 22.20 & 10.56 & 2.45 & 4.886 & 4.861 & 4.874 & 4.927 &  & 0.358 & 0.370 & 0.364 & $-$10.94 & $-$2.07 & $ 7/2^- $ \\
130 & 70 & 1065.65 & 1068.93 & 21.77 & 11.21 & 2.79 & 4.942 & 4.906 & 4.926 & 4.971 &  & 0.429 & 0.437 & 0.433 & $-$10.80 & $-$2.36 & $ 0^+ $ \\
131 & 71 & 1075.89 & 1078.17 & 21.46 & 10.25 & 2.67 & 4.972 & 4.926 & 4.951 & 4.990 &  & 0.448 & 0.454 & 0.451 & $-$10.83 & $-$2.64 & $ 7/2^- $ \\
132 & 72 & 1086.84 & 1089.90 & 21.19 & 10.95 & 2.75 & 4.991 & 4.935 & 4.965 & 4.999 & 4.917 & 0.451 & 0.459 & 0.455 & $-$11.06 & $-$3.03 & $ 0^+ $ \\
133 & 73 & 1095.56 & 1098.88 & 19.67 & 8.72 & 2.18 & 4.915 & 4.848 & 4.885 & 4.913 &  & 0.247 & 0.258 & 0.252 & $-$10.18 & $-$3.10 & $ 5/2^+ $ \\
134 & 74 & 1106.48 & 1110.26 & 19.64 & 10.92 & 2.46 & 4.923 & 4.845 & 4.888 & 4.911 & 4.913 & 0.218 & 0.233 & 0.224 & $-$10.21 & $-$3.28 & $ 0^+ $ \\
135 & 75 & 1115.74 & 1118.90 & 20.18 & 9.26 & 2.26 & 4.934 & 4.845 & 4.894 & 4.911 & 4.909 & 0.202 & 0.216 & 0.208 & $-$10.22 & $-$3.60 & $ 9/2^- $ \\
136 & 76 & 1126.35 & 1129.96 & 19.87 & 10.61 & 2.47 & 4.945 & 4.846 & 4.902 & 4.911 & 4.911 & 0.174 & 0.193 & 0.182 & $-$9.92 & $-$3.77 & $ 0^+ $ \\
137 & 77 & 1135.42 & 1138.41 & 19.68 & 9.07 & 2.18 & 4.957 & 4.847 & 4.909 & 4.912 & 4.908 & 0.153 & 0.175 & 0.162 & $-$9.92 & $-$4.04 & $ 1/2^+ $ \\
138 & 78 & 1145.92 & 1148.92 & 19.57 & 10.50 & 2.25 & 4.968 & 4.847 & 4.916 & 4.913 & 4.912 & 0.126 & 0.148 & 0.136 & $-$9.87 & $-$4.32 & $ 0^+ $ \\
139 & 79 & 1155.12 & 1156.99 & 19.70 & 9.20 & 1.57 & 4.981 & 4.850 & 4.925 & 4.916 & 4.908 & 0.114 & 0.133 & 0.122 & $-$10.26 & $-$4.66 & $ 1/2^+ $ \\
140 & 80 & 1165.77 & 1167.30 & 19.85 & 10.65 & 0.00 & 4.988 & 4.846 & 4.928 & 4.912 & 4.910 & 0.000 & 0.000 & 0.000 & $-$10.37 & $-$4.96 & $ 0^+ $ \\
141 & 81 & 1175.36 & 1175.31 & 20.24 & 9.58 & 0.00 & 5.002 & 4.851 & 4.938 & 4.917 & 4.906 & 0.039 & 0.045 & 0.042 & $-$10.52 & $-$5.29 & $ 11/2^- $ \\
142 & 82 & 1186.40 & 1185.14 & 20.63 & 11.04 & 0.00 & 5.014 & 4.854 & 4.947 & 4.920 & 4.912 & 0.000 & 0.000 & 0.000 & $-$11.28 & $-$5.56 & $ 0^+ $ \\
143 & 83 & 1191.08 & 1191.26 & 15.72 & 4.68 & 0.00 & 5.039 & 4.871 & 4.969 & 4.936 & 4.925 & $-$0.039 & $-$0.044 & $-$0.041 & $-$5.19 & $-$5.90 & $ 9/2^- $ \\
144 & 84 & 1197.43 & 1199.08 & 11.03 & 6.35 & 0.00 & 5.060 & 4.879 & 4.985 & 4.944 & 4.942 & 0.000 & 0.000 & 0.000 & $-$5.59 & $-$6.18 & $ 0^+ $ \\
145 & 85 & 1202.62 & 1204.83 & 11.55 & 5.20 & 1.06 & 5.091 & 4.902 & 5.014 & 4.967 & 4.954 & 0.125 & 0.130 & 0.127 & $-$6.10 & $-$6.59 & $ 3/2^- $ \\
146 & 86 & 1209.47 & 1212.40 & 12.05 & 6.85 & 1.86 & 5.116 & 4.915 & 5.034 & 4.979 & 4.970 & 0.152 & 0.157 & 0.154 & $-$6.46 & $-$6.88 & $ 0^+ $ \\
147 & 87 & 1215.52 & 1217.69 & 12.89 & 6.05 & 1.50 & 5.142 & 4.934 & 5.058 & 4.999 &  & 0.186 & 0.194 & 0.189 & $-$6.78 & $-$7.21 & $ 1/2^- $ \\
148 & 88 & 1222.45 & 1225.02 & 12.98 & 6.93 & 1.80 & 5.167 & 4.949 & 5.080 & 5.013 & 5.000 & 0.210 & 0.218 & 0.213 & $-$6.36 & $-$7.54 & $ 0^+ $ \\
149 & 89 & 1228.01 & 1230.06 & 12.50 & 5.57 & 1.58 & 5.197 & 4.974 & 5.108 & 5.038 &  & 0.252 & 0.266 & 0.258 & $-$6.14 & $-$8.03 & $ 1/2^+ $ \\
150 & 90 & 1235.22 & 1237.44 & 12.77 & 7.20 & 2.31 & 5.264 & 5.034 & 5.173 & 5.098 & 5.040 & 0.365 & 0.380 & 0.371 & $-$6.79 & $-$8.67 & $ 0^+ $ \\
151 & 91 & 1241.53 & 1242.77 & 13.52 & 6.32 & 2.11 & 5.284 & 5.046 & 5.191 & 5.109 &  & 0.375 & 0.387 & 0.380 & $-$6.65 & $-$10.11 & $ 11/2^- $ \\
152 & 92 & 1248.39 & 1250.05 & 13.17 & 6.85 & 2.14 & 5.289 & 5.046 & 5.194 & 5.109 &  & 0.353 & 0.370 & 0.360 & $-$6.04 & $-$10.26 & $ 0^+ $ \\
153 & 93 & 1253.57 & 1255.30 & 12.03 & 5.18 & 2.07 & 5.308 & 5.052 & 5.209 & 5.115 &  & 0.355 & 0.370 & 0.361 & $-$5.31 & $-$10.74 & $ 3/2^- $ \\
154 & 94 & 1259.34 & 1261.62 & 10.95 & 5.77 & 2.33 & 5.329 & 5.063 & 5.227 & 5.126 &  & 0.362 & 0.374 & 0.367 & $-$5.44 & $-$11.07 & $ 0^+ $ \\
155 & 95 & 1264.01 & 1266.40 & 10.45 & 4.67 & 2.22 & 5.349 & 5.073 & 5.244 & 5.135 &  & 0.370 & 0.378 & 0.373 & $-$5.33 & $-$11.45 & $ 5/2^+ $ \\
156 & 96 & 1269.85 & 1272.39 & 10.51 & 5.83 & 2.32 & 5.368 & 5.083 & 5.260 & 5.145 &  & 0.371 & 0.377 & 0.373 & $-$5.23 & $-$11.64 & $ 0^+ $ \\
157 & 97 & 1274.31 & 1276.75 & 10.29 & 4.46 & 2.17 & 5.387 & 5.092 & 5.276 & 5.154 &  & 0.378 & 0.381 & 0.379 & $-$5.04 & $-$12.00 & $ 5/2^- $ \\
158 & 98 & 1280.01 & 1282.16 & 10.16 & 5.70 & 2.17 & 5.406 & 5.102 & 5.293 & 5.164 &  & 0.379 & 0.380 & 0.379 & $-$5.03 & $-$12.19 & $ 0^+ $ \\
159 & 99 & 1284.58 & 1286.12 & 10.27 & 4.57 & 1.93 & 5.424 & 5.112 & 5.308 & 5.174 &  & 0.383 & 0.381 & 0.382 & $-$5.14 & $-$11.48 & $ 1/2^- $ \\
160 & 100 & 1289.76 & 1291.19 & 9.75 & 5.18 & 1.90 & 5.445 & 5.120 & 5.325 & 5.182 &  & 0.385 & 0.381 & 0.383 & $-$5.32 & $-$11.70 & $ 0^+ $ \\
161 & 101 & 1293.41 &  & 8.83 & 3.66 & 2.02 & 5.484 & 5.141 & 5.358 & 5.202 &  & 0.419 & 0.401 & 0.413 & $-$3.98 & $-$11.90 & $ 1/2^+ $ \\
162 & 102 & 1297.79 &  & 8.03 & 4.38 & 2.33 & 5.493 & 5.145 & 5.367 & 5.207 &  & 0.404 & 0.393 & 0.400 & $-$3.98 & $-$12.15 & $ 0^+ $ \\
163 & 103 & 1301.33 &  & 7.91 & 3.54 & 2.15 & 5.541 & 5.178 & 5.410 & 5.239 &  & 0.459 & 0.431 & 0.448 & $-$3.86 & $-$12.36 & $ 7/2^+ $ \\
164 & 104 & 1305.50 &  & 7.71 & 4.17 & 2.42 & 5.551 & 5.178 & 5.417 & 5.240 &  & 0.441 & 0.418 & 0.433 & $-$3.77 & $-$12.55 & $ 0^+ $ \\
165 & 105 & 1308.64 &  & 7.31 & 3.14 & 2.24 & 5.574 & 5.194 & 5.439 & 5.255 &  & 0.454 & 0.431 & 0.446 & $-$3.53 & $-$12.79 & $ 5/2^- $ \\
166 & 106 & 1312.56 &  & 7.06 & 3.92 & 2.42 & 5.583 & 5.192 & 5.445 & 5.253 &  & 0.433 & 0.413 & 0.426 & $-$3.49 & $-$12.97 & $ 0^+ $ \\
167 & 107 & 1315.16 &  & 6.53 & 2.60 & 2.18 & 5.548 & 5.153 & 5.409 & 5.214 &  & 0.318 & 0.310 & 0.315 & $-$3.70 & $-$13.18 & $ 1/2^- $ \\
168 & 108 & 1319.41 &  & 6.85 & 4.24 & 2.35 & 5.562 & 5.156 & 5.420 & 5.218 &  & 0.305 & 0.297 & 0.302 & $-$3.62 & $-$13.32 & $ 0^+ $ \\
169 & 109 & 1322.31 &  & 7.15 & 2.90 & 1.95 & 5.580 & 5.162 & 5.435 & 5.224 &  & 0.301 & 0.292 & 0.298 & $-$3.40 & $-$13.52 & $ 3/2^- $ \\
170 & 110 & 1326.38 &  & 6.98 & 4.07 & 2.25 & 5.593 & 5.166 & 5.446 & 5.227 &  & 0.288 & 0.280 & 0.285 & $-$3.41 & $-$13.64 & $ 0^+ $ \\
171 & 111 & 1329.22 &  & 6.92 & 2.84 & 1.84 & 5.611 & 5.176 & 5.463 & 5.237 &  & 0.288 & 0.278 & 0.285 & $-$3.13 & $-$13.89 & $ 9/2^+ $ \\
172 & 112 & 1332.72 &  & 6.34 & 3.49 & 2.36 & 5.619 & 5.174 & 5.468 & 5.235 &  & 0.264 & 0.260 & 0.262 & $-$3.10 & $-$13.94 & $ 0^+ $ \\
173 & 113 & 1334.94 &  & 5.72 & 2.22 & 2.00 & 5.631 & 5.175 & 5.477 & 5.236 &  & 0.247 & 0.248 & 0.247 & $-$2.92 & $-$14.06 & $ 9/2^+ $ \\
174 & 114 & 1338.61 &  & 5.89 & 3.67 & 2.34 & 5.641 & 5.177 & 5.486 & 5.239 &  & 0.226 & 0.231 & 0.228 & $-$3.04 & $-$14.17 & $ 0^+ $ \\
175 & 115 & 1340.70 &  & 5.76 & 2.09 & 1.69 & 5.650 & 5.171 & 5.490 & 5.232 &  & $-$0.170 & $-$0.162 & $-$0.167 & $-$3.16 & $-$14.41 & $ 7/2^+ $ \\
176 & 116 & 1344.54 &  & 5.93 & 3.84 & 2.19 & 5.663 & 5.180 & 5.503 & 5.242 &  & 0.178 & 0.190 & 0.182 & $-$3.17 & $-$14.39 & $ 0^+ $ \\
177 & 117 & 1347.05 &  & 6.35 & 2.51 & 1.77 & 5.676 & 5.179 & 5.512 & 5.240 &  & $-$0.137 & $-$0.132 & $-$0.135 & $-$3.20 & $-$14.70 & $ 1/2^- $ \\
178 & 118 & 1351.13 &  & 6.60 & 4.08 & 1.97 & 5.686 & 5.181 & 5.521 & 5.242 &  & $-$0.110 & $-$0.107 & $-$0.109 & $-$3.30 & $-$14.78 & $ 0^+ $ \\
179 & 119 & 1353.44 &  & 6.39 & 2.31 & 1.72 & 5.702 & 5.185 & 5.534 & 5.246 &  & 0.097 & 0.109 & 0.101 & $-$3.41 & $-$14.86 & $ 1/2^- $ \\
180 & 120 & 1357.64 &  & 6.51 & 4.19 & 0.00 & 5.713 & 5.186 & 5.543 & 5.247 &  & 0.041 & 0.046 & 0.043 & $-$3.51 & $-$14.93 & $ 0^+ $ \\
181 & 121 & 1360.17 &  & 6.73 & 2.54 & 1.50 & 5.726 & 5.197 & 5.557 & 5.258 &  & $-$0.057 & $-$0.059 & $-$0.058 & $-$3.34 & $-$15.20 & $ 1/2^- $ \\
182 & 122 & 1364.53 &  & 6.89 & 4.35 & 0.00 & 5.739 & 5.200 & 5.567 & 5.261 &  & 0.000 & 0.000 & 0.000 & $-$3.44 & $-$15.28 & $ 0^+ $ \\
183 & 123 & 1367.08 &  & 6.90 & 2.55 & 0.00 & 5.753 & 5.207 & 5.580 & 5.269 &  & 0.025 & 0.030 & 0.026 & $-$3.43 & $-$15.50 & $ 13/2^+ $ \\
184 & 124 & 1371.29 &  & 6.76 & 4.21 & 0.00 & 5.765 & 5.216 & 5.592 & 5.277 &  & 0.000 & 0.000 & 0.000 & $-$3.33 & $-$15.69 & $ 0^+ $ \\
185 & 125 & 1373.91 &  & 6.83 & 2.62 & 0.00 & 5.779 & 5.224 & 5.605 & 5.284 &  & 0.024 & 0.029 & 0.026 & $-$3.68 & $-$15.92 & $ 13/2^+ $ \\
186 & 126 & 1377.91 &  & 6.62 & 4.00 & 0.00 & 5.790 & 5.232 & 5.616 & 5.293 &  & 0.000 & 0.000 & 0.000 & $-$4.29 & $-$16.10 & $ 0^+ $ \\
187 & 127 & 1377.48 &  & 3.57 & \underline{$-$0.43} & 0.00 & 5.821 & 5.236 & 5.640 & 5.297 &  & 0.029 & 0.016 & 0.025 & \underline{0.17} & $-$16.23 & $ 1/2^+ $ \\
188 & 128 & 1378.45 &  & 0.54 & 0.97 & 0.00 & 5.835 & 5.246 & 5.654 & 5.306 &  & 0.000 & 0.000 & 0.000 & $-$0.36 & $-$16.38 & $ 0^+ $ \\
189 & 129 & 1377.99 &  & 0.51 & \underline{$-$0.46} & 0.00 & 5.871 & 5.249 & 5.681 & 5.310 &  & 0.041 & 0.021 & 0.035 & $-$0.35 & $-$16.50 & $ 1/2^+ $ \\
190 & 130 & 1378.95 &  & 0.50 & 0.96 & 0.00 & 5.880 & 5.259 & 5.691 & 5.320 &  & 0.000 & 0.000 & 0.000 & $-$0.34 & $-$16.65 & $ 0^+ $ \\
191 & 131 & 1378.77 &  & 0.78 & \underline{$-$0.18} & 1.63 & 5.917 & 5.273 & 5.723 & 5.333 &  & 0.128 & 0.088 & 0.115 & $-$0.63 & $-$17.01 & $ 3/2^+ $ \\
192 & 132 & 1379.99 &  & 1.04 & 1.22 & 1.32 & 5.937 & 5.284 & 5.740 & 5.344 &  & 0.141 & 0.100 & 0.128 & $-$0.74 & $-$17.21 & $ 0^+ $ \\
193 & 133 & 1380.09 &  & 1.33 & 0.11 & 1.43 & 5.962 & 5.299 & 5.764 & 5.359 &  & 0.173 & 0.126 & 0.158 & $-$0.83 & $-$17.46 & $ 3/2^+ $ \\
194 & 134 & 1381.34 &  & 1.36 & 1.25 & 1.42 & 5.981 & 5.305 & 5.781 & 5.365 &  & 0.174 & 0.129 & 0.160 & $-$0.74 & $-$17.56 & $ 0^+ $ \\
195 & 135 & 1381.45 &  & 1.35 & 0.10 & 1.29 & 6.011 & 5.314 & 5.806 & 5.374 &  & 0.194 & 0.142 & 0.178 & $-$0.77 & $-$17.69 & $ 1/2^+ $ \\
196 & 136 & 1382.68 &  & 1.33 & 1.23 & 1.54 & 6.026 & 5.325 & 5.820 & 5.385 &  & 0.200 & 0.153 & 0.186 & $-$0.74 & $-$17.84 & $ 0^+ $ \\
197 & 137 & 1382.79 &  & 1.35 & 0.11 & 1.67 & 6.051 & 5.335 & 5.842 & 5.395 &  & 0.214 & 0.165 & 0.199 & $-$0.66 & $-$17.96 & $ 1/2^+ $ \\
198 & 138 & 1384.01 &  & 1.33 & 1.21 & 1.56 & 6.070 & 5.345 & 5.860 & 5.405 &  & 0.222 & 0.175 & 0.208 & $-$0.70 & $-$18.07 & $ 0^+ $ \\
199 & 139 & 1383.98 &  & 1.19 & \underline{$-$0.03} & 1.67 & 6.093 & 5.355 & 5.881 & 5.415 &  & 0.234 & 0.185 & 0.219 & $-$0.66 & $-$18.19 & $ 5/2^+ $ \\
200 & 140 & 1385.29 &  & 1.29 & 1.31 & 1.80 & 6.144 & 5.388 & 5.928 & 5.447 &  & $-$0.255 & $-$0.238 & $-$0.250 & $-$0.94 & $-$18.57 & $ 0^+ $ \\
201 & 141 & 1385.36 &  & 1.39 & 0.07 & 1.75 & 6.171 & 5.394 & 5.949 & 5.453 &  & $-$0.259 & $-$0.241 & $-$0.254 & $-$0.95 & $-$18.69 & $ 3/2^+ $ \\
202 & 142 & 1386.78 &  & 1.49 & 1.42 & 1.93 & 6.180 & 5.406 & 5.961 & 5.465 &  & $-$0.260 & $-$0.242 & $-$0.255 & $-$0.72 & $-$18.87 & $ 0^+ $ \\
203 & 143 & 1386.76 &  & 1.40 & \underline{$-$0.02} & 1.85 & 6.204 & 5.412 & 5.981 & 5.471 &  & $-$0.263 & $-$0.243 & $-$0.257 & $-$0.68 & $-$18.98 & $ 3/2^+ $ \\
204 & 144 & 1387.98 &  & 1.19 & 1.22 & 2.01 & 6.216 & 5.422 & 5.993 & 5.481 &  & $-$0.263 & $-$0.242 & $-$0.257 & $-$0.63 & $-$19.14 & $ 0^+ $ \\
205 & 145 & 1387.88 &  & 1.12 & \underline{$-$0.10} & 1.86 & 6.238 & 5.429 & 6.013 & 5.488 &  & $-$0.266 & $-$0.243 & $-$0.259 & $-$0.58 & $-$19.27 & $ 3/2^+ $ \\
206 & 146 & 1389.04 &  & 1.07 & 1.16 & 2.01 & 6.251 & 5.437 & 6.025 & 5.496 &  & $-$0.266 & $-$0.242 & $-$0.259 & $-$0.57 & $-$19.41 & $ 0^+ $ \\
207 & 147 & 1388.89 &  & 1.01 & \underline{$-$0.16} & 2.08 & 6.273 & 5.444 & 6.044 & 5.503 &  & $-$0.268 & $-$0.243 & $-$0.261 & $-$0.51 & $-$19.54 & $ 1/2^+ $ \\
208 & 148 & 1389.99 &  & 0.95 & 1.10 & 1.92 & 6.287 & 5.452 & 6.058 & 5.510 &  & $-$0.269 & $-$0.242 & $-$0.261 & $-$0.49 & $-$19.67 & $ 0^+ $ \\
209 & 149 & 1389.68 &  & 0.79 & \underline{$-$0.31} & 1.94 & 6.307 & 5.461 & 6.076 & 5.519 &  & $-$0.271 & $-$0.244 & $-$0.263 & $-$0.35 & $-$19.82 & $ 1/2^+ $ \\
210 & 150 & 1390.74 &  & 0.75 & 1.07 & 1.80 & 6.322 & 5.468 & 6.090 & 5.526 &  & $-$0.271 & $-$0.243 & $-$0.263 & $-$0.35 & $-$19.94 & $ 0^+ $ \\
211 & 151 & 1390.34 &  & 0.67 & \underline{$-$0.40} & 1.47 & 6.343 & 5.479 & 6.109 & 5.537 &  & $-$0.278 & $-$0.249 & $-$0.270 & $-$0.28 & $-$20.11 & $ 13/2^- $ \\
212 & 152 & 1391.16 &  & 0.42 & 0.82 & 1.80 & 6.357 & 5.484 & 6.122 & 5.542 &  & $-$0.271 & $-$0.243 & $-$0.263 & $-$0.18 & $-$20.20 & $ 0^+ $ \\
213 & 153 & 1390.67 &  & 0.32 & \underline{$-$0.49} & 1.92 & 6.387 & 5.487 & 6.147 & 5.545 &  & $-$0.266 & $-$0.241 & $-$0.259 & $-$0.15 & $-$20.28 & $ 1/2^+ $ \\
214 & 154 & 1391.26 &  & 0.10 & 0.59 & 1.87 & 6.390 & 5.495 & 6.152 & 5.553 &  & $-$0.264 & $-$0.237 & $-$0.257 & $-$0.07 & $-$20.43 & $ 0^+ $ \\
215 & 155 & 1390.68 &  & 0.02 & \underline{$-$0.58} & 1.76 & 6.417 & 5.500 & 6.175 & 5.558 &  & $-$0.258 & $-$0.235 & $-$0.252 & $-$0.02 & $-$20.52 & $ 1/2^+ $ \\
216 & 156 & 1391.20 &  & \underline{$-$0.06} & 0.52 & 1.89 & 6.421 & 5.503 & 6.179 & 5.561 &  & $-$0.252 & $-$0.225 & $-$0.244 & $-$0.03 & $-$20.62 & $ 0^+ $ \\
\hline 

\end{longtable}

\newpage

\end{landscape}

\end{document}